\documentclass[aip,amsmath,amssymb,graphicx,reprint]{revtex4-2}

\usepackage{graphicx}
\usepackage{graphicx}
\usepackage{color}
\usepackage{bm}
\usepackage{hyperref} 
\usepackage{cancel}
\usepackage{dcolumn}
\usepackage{xcolor}
\usepackage{booktabs}
\usepackage{tabularray}
\usepackage{booktabs}
\usepackage{longtable}
\usepackage{float}

\graphicspath{ {./Images/} }

\usepackage[utf8]{inputenc}
\usepackage[T2A]{fontenc}
\usepackage[normalem]{ulem} 

\draft 

\begin{document}

\title{
Impact of ligand (OH) deformation on LuOH$^+$ rovibrational spectra 
} 

\author{Igor Kurchavov} \email{kurchavov\_ip@pnpi.nrcki.ru}
\author {Sergey Prosnyak}\email{prosnyak\_sd@pnpi.nrcki.ru}
\affiliation{Petersburg Nuclear Physics Institute named by B.P. Konstantinov of National Research Centre
"Kurchatov Institute", Gatchina, 1, mkr. Orlova roshcha, 188300, Russia}

\author {Leonid V. Skripnikov} \email{skripnikov\_lv@pnpi.nrcki.ru}
\author {Alexander Petrov} \email{petrov\_an@pnpi.nrcki.ru}

\affiliation{Petersburg Nuclear Physics Institute named by B.P. Konstantinov of National Research Centre
"Kurchatov Institute", Gatchina, 1, mkr. Orlova roshcha, 188300, Russia}
\affiliation{St. Petersburg State University, St. Petersburg, 7/9 Universitetskaya nab., 199034, Russia}

\date{\today}

\begin{abstract}

Triatomic cation $^{175}$LuOH$^+$, featuring near-degenerate, opposite-parity $l$-doublets, offers enhanced sensitivity to $\mathcal{P}$- and $\mathcal{T}$-violating interactions. We present \emph{ab initio} calculations of its electronic structure and rovibrational structure beyond the rigid-ligand approximation by explicitly including OH-ligand deformation together with bending and stretching motions. Potential-energy surfaces are computed at the relativistic coupled cluster level of theory. The nuclear Schr\"{o}dinger equation in Jacobi coordinates is solved by means of a coupled-channel expansion. Ligand deformation reduces the bending frequency by a few percent and increases the $l$-doubling constant $q$, while the stretching frequencies and rotational constants remain largely unchanged. For the first excited bending level, we predict $\Delta E_{J=1}=2q \approx 24.9$--$26.4$ MHz. These results establish LuOH$^+$ as a viable platform for precision searches for $\mathcal{CP}$-violating physics via the electron electric dipole moment and the nuclear magnetic quadrupole moment.

\end{abstract}


\maketitle 
\section{Introduction}

Parity ($\mathcal{P}$), time reversal ($\mathcal{T}$), and charge conjugation ($\mathcal{C}$) symmetries are violated in the Standard Model (SM)~\cite{schwartz2014quantum,particle2020review,khriplovich2012cp}. Combined $\mathcal{CP}$ nonconservation arises in the SM weak interaction when a complex phase appears in the Cabibbo-Kobayashi-Maskawa (CKM) \cite{Cabibbo1963,KobayashiMaskawa1973} matrix, associated with quark–$W^\pm$ interactions, or in the Pontecorvo–Maki–Nakagawa–Sakata (PMNS)~\cite{Pontecorvo1957,MNS1962} matrix, associated with lepton–$W^\pm$ interactions. Direct violation of $\mathcal{CP}$ has been observed in neutral kaon and $B^0$ -meson decays \cite{schwartz2014quantum} and in charmed $D^0$-meson decays \cite{CharmDecay19}. However, within the SM the magnitude of $\mathcal{CP}$ violation is insufficient to explain the baryon asymmetry of the Universe. $\mathcal{CP}$ violation in the SM also gives rise, in principle, to an electron electric dipole moment ($e$EDM) ($d_e$) and a scalar–pseudoscalar interaction between electrons and nucleons (S-PS) ($k_s$) \cite{PospelovRitz2014,YamaguchiYamanaka2020}. Many extensions of the SM predict much larger values of $d_e$ and $k_s$ \cite{Fukuyama2012, YamaguchiYamanaka2021}. Thus, measurements of the $e$EDM serve as a highly sensitive probe of the SM and its extensions \cite{whitepaper, YamaguchiYamanaka2020,YamaguchiYamanaka2021}. To date, the most stringent experiments have used linear diatomic molecules to probe $\mathcal{CP}$-violating physics. Recently, the JILA group obtained a new constraint on the $e$EDM, $|d_e|<4.1\times 10^{-30}$ $e \cdot \mathrm{cm}\ $ (90\% confidence) \cite{JILA23}, using $^{180}$Hf$^{19}$F$^+$ ions trapped in a rotating electric field.

One possible source of $\mathcal{CP}$ violation in the Standard Model is the $\theta$-term in the QCD Lagrangian. Experimental searches for the neutron electric dipole moment constrain the effective angle $\bar\theta$ to be extremely small, with upper limits ranging from $\bar\theta \lesssim 10^{-10}$ \cite{KimCarosi2010} to the more recent bound of $|\bar\theta| < 1.5\times 10^{-8}$ (90 \% confidence) based on paramagnetic molecular experiments with HfF$^+$ \cite{Mulder2025}, giving rise to the strong $\mathcal{CP}$ problem. One possible solution is to introduce a continuous dynamical field whose spontaneous breaking leads to a pseudo-Goldstone boson (the axion), which dynamically suppresses the effective $\theta$ angle.

In turn, $^{175}$Lu nucleus with spin  $I=5/2$ acquires a $\mathcal{CP}$-odd magnetic quadrupole moment (MQM). The MQM can be expressed in terms of the proton EDM, the QCD vacuum angle $\theta$, and the quark chromo-EDMs. Measurements of nuclear MQMs are therefore promising for establishing new constraints on $\mathcal{CP}$ violation in the hadronic sector.

Cold polar  molecular systems provide unique opportunities to probe the effects of symmetry violation \cite{Isaev:16}. Among them, triatomic systems such as LuOH$^+$~\cite{Maison:20a}, YbOH, and their analogs exhibit a rovibrational structure with additional parity doublets ($l$-doublets), enhancing sensitivity to such effects \cite{Kozyryev:17, hutzler2020polyatomic, Petrov:2022, Petrov:24b}. In Ref.~\cite{Petrov:2022, Petrov:24a} we showed that the magnitude of the $l$-doubling and the spin–rotation splitting directly determines the maximum achievable $\mathcal{T},\mathcal{P}$-odd polarization and thus the sensitivity of linear triatomic molecules to $\mathcal{T},\mathcal{P}$-odd effects.

In our previous work \cite{zakharova22}, we calculated the $l$-doubling in YbOH while explicitly including ligand deformation as well as bending and stretching vibrations, highlighting the importance of these effects for accurate predictions. Subsequent measurements of the $l$-doubling in YbOH \cite{jadbabaie23} confirmed these predictions, particularly when OH deformation is taken into account. This agreement underscores the critical role of vibrational effects in theoretical studies of related molecules. Motivated by these results, we now apply this approach to the $^{175}$LuOH$^+$ cation, for which our most recent work considered only bending and stretching modes \cite{kurchavov23}.
In addition to this, we have investigated these results using larger basis sets.

Our work focuses on calculating the electronic structure, vibrational modes, and $l$-doubling characteristics of $^{175}$LuOH$^+$. To this end, we (i) construct potential-energy surfaces on a $(R,r,\theta)$ grid; (ii) interpolate the surfaces using constrained multivariate polynomial regression; and (iii) solve the nuclear Schr\"{o}dinger equation in Jacobi coordinates within the Born–Oppenheimer approximation via a coupled-channel expansion over OH vibrational functions, from which we extract vibrational/rotational constants and the $l$-doubling parameter $q$. Particular attention is given to the role of OH-ligand deformation. These results establish the suitability of LuOH$^+$ as a candidate for precision symmetry-violation experiments.

\section{Methods}
\subsection{Rovibrational energy calculations}
We use the Born–Oppenheimer approximation to factorize the total wavefunction into $\Psi_{\rm nuc}$, which describes the nuclear motion in the adiabatic potential of the electrons, and $\Psi_{\rm elec}$, which corresponds to the electronic motion in the field of the nuclei,
\begin{equation}
\Psi_{\rm total}\approx\Psi_{\rm nuc}(\vec{R}, \vec{r})\psi_{\rm elec}(\vec{R}, \vec{r}|Q),
\label{totalWF}
\end{equation}
where $Q$ denotes the generalized electronic coordinates, $\vec{R}$ is the vector from the lutetium atom to the center of mass of the OH ligand, and $\vec{r}$ is the vector from the oxygen to the hydrogen atom.

\begin{figure}[h]
\centering
  \includegraphics[width=0.4\textwidth]{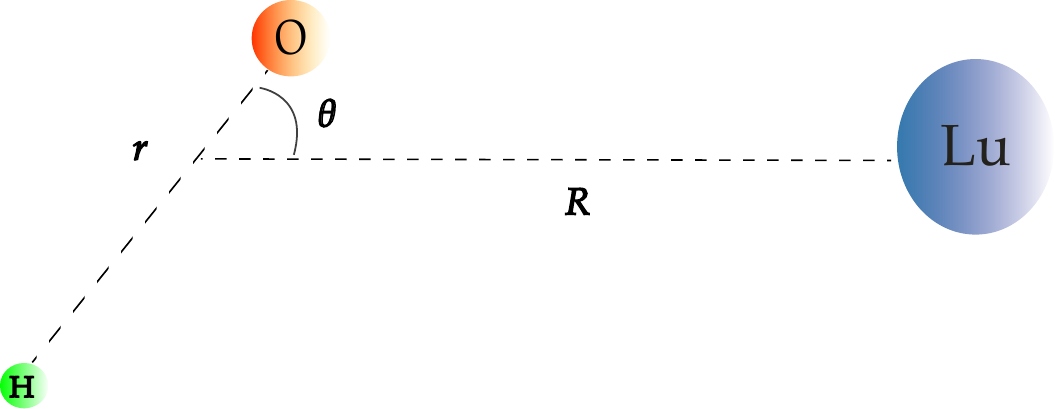}
  \caption{The Jacobi coordinates for the LuOH$^+$ cation.}
  \label{Jacob}
\end{figure}

The nuclear Hamiltonian has the form
\begin{equation}
\begin{aligned}
    \hat{H}_{\rm nuc}=-\frac{1}{2\mu}\frac{\partial^2}{\partial R^2} -\frac{1}{2\mu_{OH}}\frac{\partial^2}{\partial r^2}+\frac{\hat{L}^2}{2\mu R^2}+\frac{\hat{j}^2}{2\mu_{OH}r^2}\\+V(R, r,\theta),
\end{aligned}
\end{equation}
where $\theta$ is the angle between the vectors $\vec{R}$ and $\vec{r}$ (so that $\theta=0$ corresponds to the Lu-O-H linear configuration); $\mu$ is the Lu-OH reduced mass; $\mu_{OH}$ is the reduced mass of the ligand; $\hat{L}$ and $\hat{j}$ are rotational angular momenta
of Lu and OH around their center of mass and OH, respectively. The directions of the axes $\hat{r}$ and $\hat{R}$ are shown in Fig.~\ref{Jacob}. $V$~is the electronic potential-energy surface obtained at the CCSD(T) level and depends only on the relative Jacobi coordinates $(R,r,\theta)$.

The nuclear wavefunction $\Psi_{\rm nuc}(\vec{R}, \vec{r})$ is the solution of the Schr\"{o}dinger equation
\begin{equation}
\hat{H}_{nuc}\Psi_{\rm nuc}(\vec{R}, \vec{r}) = E \Psi_{\rm nuc}(\vec{R}, \vec{r}).
\label{Shreq}
\end{equation}
To solve Eq. (\ref{Shreq}) we use the expansion
\begin{equation}
\Psi_{\rm nuc}(\vec{R}, \vec{r}) = \sum_{L=0}^{L_{max}}\sum_{j=0}^{j_{max}}\sum_{n=1}^{n_{max}} F_{JjLn}(R)\Phi_{JjLM}(\hat{R},\hat{r})f_n(r),
\label{psiexp}
\end{equation}
where
\begin{equation}
\Phi_{JjLM}(\hat{R},\hat{r}) = \sum_{m_L,m_j} C^{JM}_{Lm_L,jm_j} Y_{Lm_L}(\hat{R})Y_{jm_j}(\hat{r})
\end{equation}
is coupled to conserved total angular momentum $J$ basis set, $Y_{Lm_L}$ is a spherical function, $f_n(r)$ is a solution of
\begin{equation}
 \left(-\frac{1}{2\mu_{OH}}\frac{\partial^2}{\partial r^2}+V(R_i, r,\theta_i)\right)f_n(r) = e_n f_n(r),
 \label{wOH}
\end{equation}
where $R_i$ and $\theta_i$ are some fixed values. 
(For simplicity we do not indicate the dependence
of $f_n(r)$ on $R_i$ and $\theta_i$)
In Ref. \cite{zakharova22} we tested that the final result is independent of the choice of  $(R_i,\theta_i)$ values.

Substituting the wavefunction (\ref{psiexp}) into Eq.~(\ref{Shreq}), we obtain a system of close-coupled equations for
$F_{JjLn}(R)$ \cite{mcguire1974quantum}.

\subsection{Electronic structure}
In our previous paper~\cite{kurchavov23}, to construct the potential-energy surface of the molecular cation LuOH$^+$, we used the four-component Dirac–Coulomb Hamiltonian and a basis set denoted here as PBas. It is based on Dyall’s uncontracted all-electron triple-zeta AE3Z basis set~\cite{gomes2010relativistic} for Lu and the aug-cc-pVTZ-DK basis sets~\cite{Dunning:89, Kendall:92} for O and H. Our analysis indicates that electron spin–dependent relativistic effects make only a modest contribution to the results. This allows us to substantially increase the basis-set size in {\it one-component} (scalar-relativistic) calculations and thereby reduce the associated uncertainties. We achieved this by using the valence part of the 28-electron generalized relativistic effective core potential (GRECP) developed by the PNPI Quantum Physics and Chemistry Department~\cite{titov1999generalized,mosyagin2010shape,mosyagin2016generalized}, with the spin–orbit term removed. Accordingly, the leading approximation for potential-energy surface was obtained using the large HBas basis set, based on Dyall’s all-electron quadruple-zeta AE4Z basis for Lu~\cite{gomes2010relativistic} and the aug-cc-pVQZ-DK basis sets for O and H~\cite{Dunning:89, Kendall:92}. To account for spin–orbit effects, we computed a correction defined as the difference between results obtained with and without the inclusion of spin–orbit coupling. Due to the high cost of two-component calculations, we used the smaller and compact SBas basis set to compute this correction. The SBas basis set was constructed following the compact-basis-generation procedure~Refs.\cite{Skripnikov:2020e,Skripnikov:13a} applied to the Lu atom starting from the AETZ basis set~\cite{gomes2010relativistic} as the initial approximation. The SBas for O and H atoms corresponds to the aug-cc-pVTZ-DK basis set~\cite{Dunning:89, Kendall:92}.
Additionally, we employed the MBas basis, obtained by removing the tightest $s$- and $p$-type primitive Gaussians on Lu from PBas, since these functions contribute negligibly to the calculated properties owing to the smooth behavior of the valence wave function in the core region within the GRECP framework~\cite{titov1999generalized,mosyagin2010shape,mosyagin2016generalized}. 
The compositions of all basis sets are given in Table~\ref{tab:bsc}. In all calculations, virtual-orbital energies were truncated at 300 a.u.

Electron correlation was treated using the coupled cluster method with single and double excitations augmented by perturbative triple excitations, CCSD(T)~\cite{Bartlett:2007}.
Two-component (spin–orbit) calculations with the SBas basis set were carried out using {\sc dirac}~\cite{DIRAC19,Saue:2020}, whereas one-component scalar-relativistic calculations with the SBas, MBas, and HBas basis sets were performed using {\sc cfour}\cite{CFOUR}. The {\sc natbas} code~\cite{Skripnikov:2020e,Skripnikov:13a} was employed to construct the SBas basis set.

\begin{table}
    \centering
    \caption{Composition of basis sets employed in calculations. In parenthesis we present the number of uncontracted gaussian functions, while in square brackets the number of contracted ones.}
    \label{tab:bsc}
\begin{tblr}{cccl}
        \hline
        \hline
         &  Lu& O&H\\
         \hline
         4c/PBas&  { (30s24p16d12f6g2h) \\ $[$30s24p16d12f6g2h$]$ }&  { (11s6p3d2f) \\ $[$5s4p3d2f$]$ }&{ (6s3p2d) \\ $[$4s3p2d$]$ }\\
         1c/SBas&  { (19s17p13d10f6g) \\ $[$19s17p13d6f3g$]$ }&  { (11s6p3d2f) \\ $[$5s4p3d2f$]$ }&{ (6s3p2d) \\ $[$4s3p2d$]$ }\\
         1c/MBas&   { (23s20p16d12f6g2h) \\ $[$23s20p16d12f6g2h$]$} &  { (11s6p3d2f) \\ $[$5s4p3d2f$]$ }&{ (6s3p2d) \\ $[$4s3p2d$]$ }\\
         1c/HBas&  { (32s27p24d15f10g2h) \\ $[$32s27p24d15f10g2h$]$ }&  { (13s7p4d3f2g) \\ $[$6s5p4d3f2g$]$ }&{ (7s4p3d2f) \\ $[$5s4p3d2f$]$ }\\
        \hline
        \hline
    \label{tab:my_label}
\end{tblr}
\end{table}

\subsection{Potential surface interpolation}
We calculated potential-energy surfaces, denoted \(V\), at the 
CCSD(T) level using various basis sets for a grid of coordinates \( (R_i, r_j, \theta_k) \):
\[
\begin{aligned}
\{R_i\} &= 1.771,           1.877,           1.930,           1.983,           2.089 \text{\AA} \\
\{r_j\} &= 0.854, 0.904, 0.954, 1.004, 1.054 \text{\AA} \\
\{\theta_k\} &= 0^\circ, 5^\circ, 10^\circ, 15^\circ, 20^\circ, 25^\circ, 55^\circ, 90^\circ, 122^\circ, 155^\circ, 180^\circ
\end{aligned}
\]
For potential-surface interpolation, we use a three-variable multivariate polynomial regression model, expressed as:

\[
V(R, \theta, r) = \sum_{i=0}^{d_1} \sum_{j=0 \atop j \ne 1}^{d_2} \sum_{k=0}^{d_3} C_{ijk} R^i \theta^j r^k + \varepsilon,
\]
where 
\( R, \theta, r \) are independent variables, \( C_{ijk} \) represents the polynomial coefficients, and \( \varepsilon \) denotes the residual error. The maximum polynomial degrees \( d_1, d_2, d_3 \) define the complexity of the regression model. For $\theta$ we neglected the linear term, so that the derivative becomes zero when $\theta = 0$:
\[
\left. \frac{\partial V}{\partial \theta} \right|_{\theta = 0} = 0.
\]

The accuracy of the regression model can be assessed using 
root mean square error ($\varepsilon_\mathrm{RMSE}$), 
$$
\varepsilon_{\mathrm{RMSE}} = 
\sqrt{ \frac{ \displaystyle\sum_{i=1}^{N} w_i \left( V_i - \hat{V}_i \right)^2 }
{\displaystyle\sum_{i=1}^{N} w_i } }
$$
where
$V_i$ — observed calculated value, $\hat{V}_i$ — predicted value from the model, $w_i$ — weight of the $i$-th point, $N$ — total number of points.
$\varepsilon_\mathrm{RMSE}$ emphasizes larger deviations due to the squared term, making it more sensitive to significant errors. Weighted fitting was employed by assigning larger weights to the first six bending angles and using the averaged weights for the remaining coordinates. This strategy prioritizes the physically most relevant part of the potential energy surface while preventing overfitting.

Table \ref{tab:rmse} shows the obtained $\varepsilon_\mathrm{RMSE}$ values for different basis sets, corresponding to the optimized polynomial degrees of $d_1=4$, $d_2=10$ and $d_3 = 4$. 

\begin{table}
    \centering
    \caption{Root mean square error ($\varepsilon_\mathrm{RMSE}$) in a.u., for different basis sets. ``+OH'' and ``+SO'' indicate inclusion of ligand deformation and spin–orbit effects, respectively.  $\varepsilon_\mathrm{RMSE}(20)$ means for those energies within $20^\circ$. ``4c'' denotes a four-component Dirac–Coulomb calculation, whereas ``1c'' designates the scalar-relativistic approximation applied to the outer-core and valence electrons.}
    \begin{tblr}{l|cc}
    \hline \hline
         & $\varepsilon_\mathrm{RMSE}$ & $\varepsilon_\mathrm{RMSE}(20)$\\ \hline
      4c/PBas     &   $1.14\times10^{-5}$   &   $1.85\times10^{-6}$     \\
      1c/MBas     &   $1.95\times10^{-7}$   &   $3.76\times10^{-8}$     \\
      1c/HBas     &   $2.77\times10^{-7}$   &   $4.86\times10^{-8}$     \\
  1c/HBas +OH     &   $2.06\times10^{-7}$   &   $2.02\times10^{-7}$     \\
1c/HBas +OH +SO   &   $2.03\times10^{-7}$   &   $2.01\times10^{-7}$     \\ \hline \hline
    \end{tblr}
    \label{tab:rmse}
\end{table}

\begin{figure}
    \centering
    \includegraphics[width=1\linewidth]{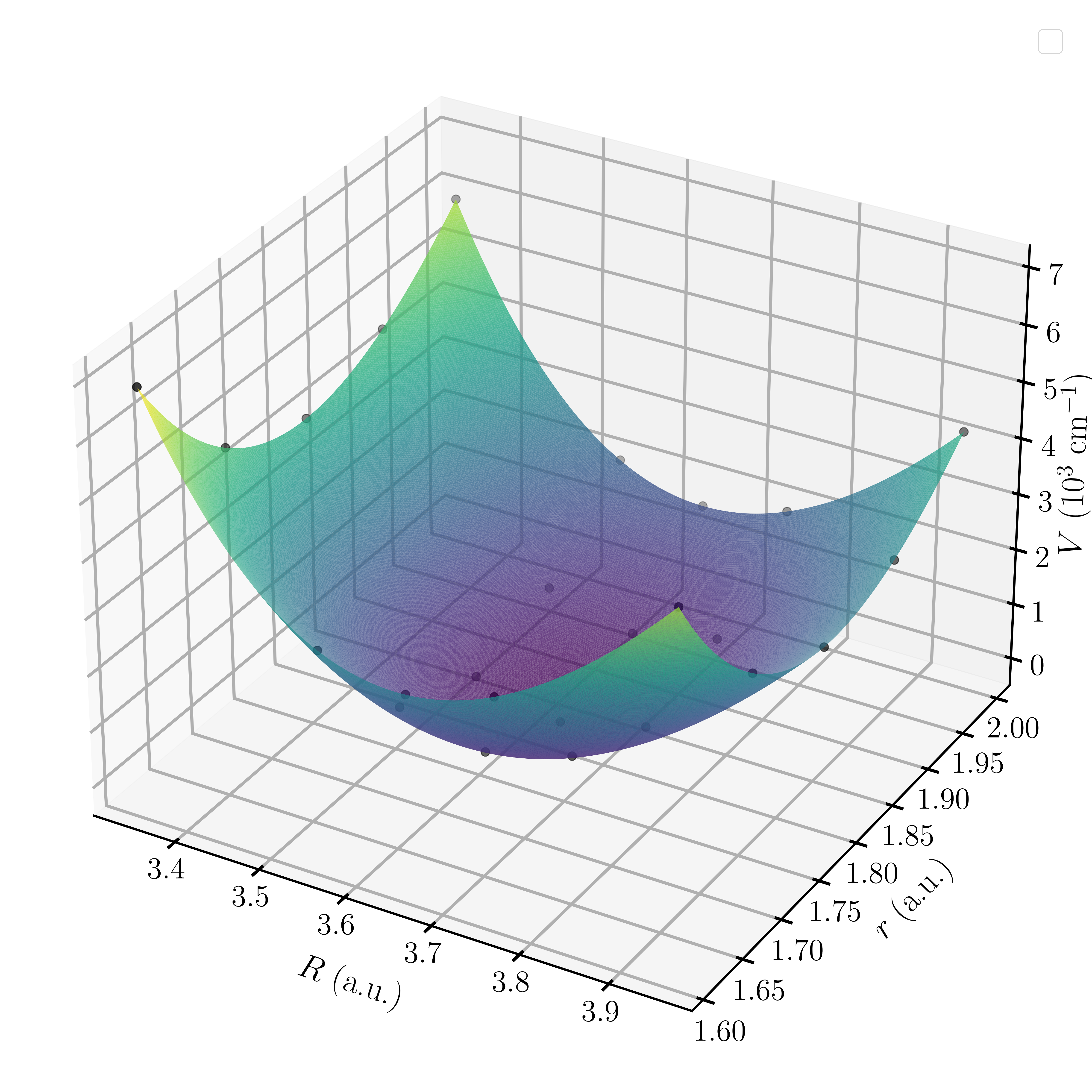}
    \caption{The two-dimensional projection of the potential energy surface was calculated using the HBas basis set, taking into account spin-orbit, and corresponds to a fixed $\theta=0$.}
    \label{fig:PES}
\end{figure}
\begin{figure}
    \centering
    \includegraphics[width=1\linewidth]{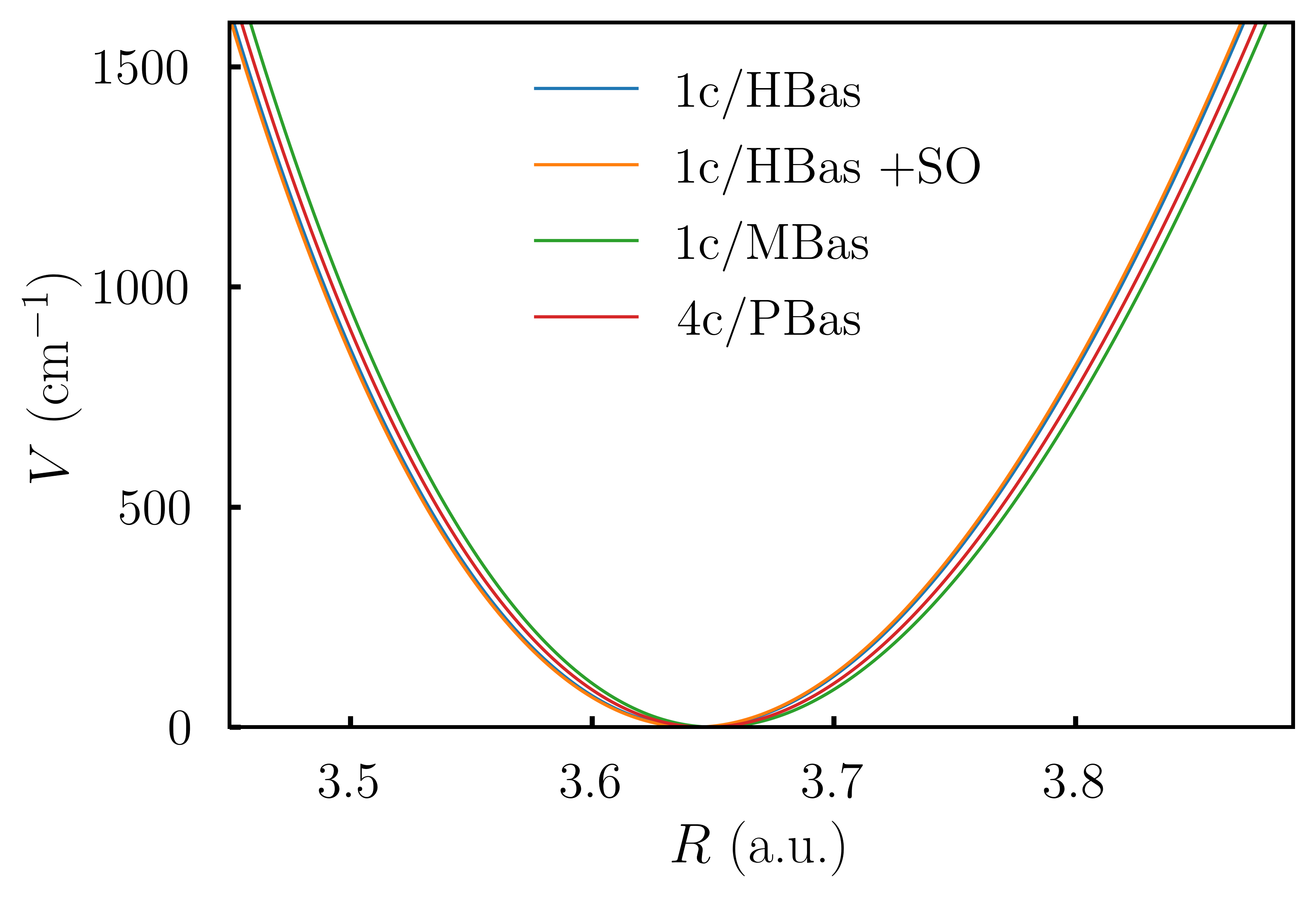}
    \caption{The one-dimensional projections on $R$ variable of the calculated potential energy surfaces for different basis sets.}
    \label{fig:bases_R}
\end{figure}
\begin{figure}
    \centering
    \includegraphics[width=1\linewidth]{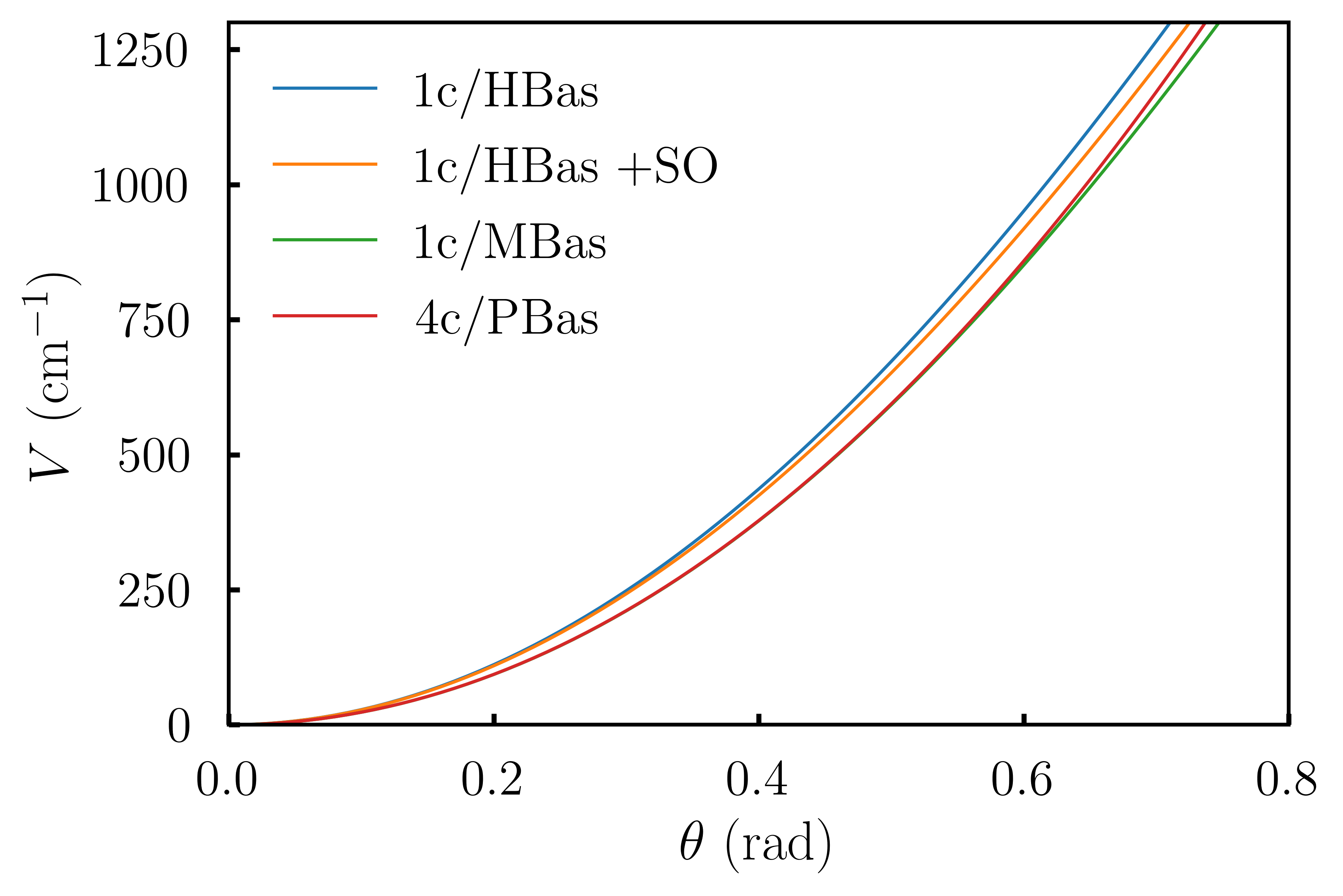}
    \caption{
    The one-dimensional projections onto the $\theta$ variable of the calculated potential energy surfaces for different basis sets.}
    \label{fig:bases_theta}
\end{figure}
\section{Results}
The two-dimensional projection of the potential energy surface calculated with the largest basis set and taking into account the spin-orbit interaction is shown in Fig. \ref{fig:PES}.
The interpolation scheme employed in this work demonstrated stability, allowing us to reproduce the $V(R,\theta,r)$ features with satisfactory accuracy. 
For most basis/method combinations, the fit yields small residuals and stable rovibrational constants; however, for the 4c/Pbas data set, the achieved accuracy is somewhat lower (see Table \ref{tab:rmse}). 
For this basis set, we performed new calculations using regression.
We attribute this degraded performance primarily to the limited numerical precision in the input energies for that basis (fewer significant digits), which reduces the information available to the fitting procedure. 

Table~\ref{tab:VibSpectrum} and Table~\ref{tab:RotSpectrum} show the calculated vibrational and rotational energies, respectively. The influences of increasing the basis set, spin-orbit interaction, and inclusion of the OH deformation are shown. The
vibrational states are assigned using the notations ($\nu_1$,$\nu_2^l$,$\nu_3$), where $\nu_1$, $\nu_2$, and $\nu_3$ are vibrational quantum numbers for the modes associated with $R$ (stretching mode), $\theta$ (bending mode) and $r$ (OH stretching), respectively.
$l-$ is the absolute value of the projection of the vibrational angular momentum (of the bending mode) on $R$ bond.
$l$ takes the values $ l = \nu_2, \nu_2-2, \cdots,1(0)$.
We note, that our rovibrational structure calculations go beyond the harmonic approximations and all modes interact with each other. This can be seen, for example, from the fact that the inclusion of the OH ligand deformation influences the energies of other modes.

One can see that taking into account the ligand deformation decreases the vibrational energy of the first bending mode for the HBas basis set by about 2.4\% percent, and the spin-orbit interaction decreases it further by about 1.5\% percent while the $l$-doubling 
is increasing. 
The largest effect is observed  when increasing the basis set from Mbas to Hbas. The energy of the first bending mode increases in this case by about 5.9\%, while the $l$-doubling decreases.
The decrease (increase) of the $l$-doubling value together with the increase (decrease) of the $\nu_2$ frequency is consistent with the estimate \cite{HerzbergBook}
\begin{equation}
q \simeq 
    \frac{B^2}{\nu_2}\Big(1+4\sum_{k=1,3}\frac{\zeta_{2k}^2\nu_{2}^2}{\nu_{k}^2-\nu_{2}^2}\Big)(v+1),\quad \Delta E=2q,
\end{equation}
where $\zeta_{2k}$ are Coriolis coefficients.
The stretching vibrational energies remains largely unchanged across various basis sets,
and accounting spin-orbit and ligand deformation effects

The one-dimensional projections on $R$ and $\theta$ variables of the calculated potential energy surfaces are given on Fig. \ref{fig:bases_R} and Fig. \ref{fig:bases_theta} respectively.
The potential energy surfaces calculated with the Pbas basis set (four-component calculation \cite{Maison:22, kurchavov23} ), Mbas basis set, Hbas basis set, and Hbas basis set with spin-orbit effect included are given.
Fig. \ref{fig:bases_R} and Fig. \ref{fig:bases_theta} give qualitative clue on the extent of the influence of the basis set increasing and spin-orbit effect on the stretching and bending modes respectively. One can see that projections on $R$ variable are much closer to each other than projections on $\theta$ variable.

In Table~\ref{tab:Conv}, one can see that the results converge as the quantum number $n_{max}$ (see Eq.~(\ref{psiexp})) increases.
Table~\ref{tab:Conv} also shows that for purposes of the present paper it is sufficient to use $n_{max}=2$.

From Table~\ref{tab:VibSpectrum}, we estimate the accuracy of our calculations for the $l$-doubling of the first bending mode to be at the level of 3\%. The inclusion of the iterative triple excitations to the coupled cluster computations and using more extensive basis sets are necessary to reach the higher accuracy. Such calculations, however, are very demanding in terms of computational resources and are therefore beyond the scope of the present work.  For the second excited bending mode, the $l$-doubling effect is found to be very small, and the calculated value should be considered only as an order-of-magnitude estimate.

Overall, our study indicates that the first excited bending $v_2=1$ $l$-doublets in LuOH$^+$ is exceptionally favorable for $\mathcal{CP}$-violation searches providing nearly degenerate parity states that can be fully mixed by modest laboratory fields, enabling long coherence times.

\begin{table*}
\label{tab:Conv} 
\caption{Convergence of the results with $n_{max}$ (see Eq. (\ref{psiexp})) for different basis sets for CCSD(T) calculations. ``+OH'' and ``+SO'' are associated with ligand deformation and spin-orbit accordingly.}
\begin{ruledtabular}
\begin{tabular}{clccc}
parameter                             & Basis set              &   $n_{max}=1$ & $n_{max}=2$ & $n_{max}=3$ \\ \hline
Bending mode $\nu_2$, cm$^{-1}$       &     1c/HBas +OH                 &  450.68     &  449.13   &  449.11   \\
                                      &     1c/HBas +OH +SO             &  443.62     &  442.25   &  442.23 \\
$l$-doubling $\Delta E_{J=1}=2q$, MHz &     1c/HBas +OH                 &  24.04       & 25.26    &  25.27    \\
                                      &     1c/HBas +OH +SO              &  24.51       & 25.71    &  25.72
\end{tabular}
\end{ruledtabular}
\end{table*}
\section{Conclusion} 
New three-dimensional potential energy surfaces using several basis sets are constructed for the ground state of the LuOH$^+$ cation. The lowest rovibrational energy levels
and $l$-doubling for the first and second bending modes are calculated.
Influence of basis set increasing, spin-orbit interaction and OH ligand deformation are investigated.
We expect that our results will be useful in further experimental and theoretical studies of $\mathcal{P}\mathcal{T}$-odd properties in  LuOH$^+$.

\begin{acknowledgments}
Electronic structure calculations have been carried out using computing resources of the federal collective usage center Complex for Simulation and Data Processing for Mega-science Facilities at National Research Centre ``Kurchatov Institute'', \url{http://ckp.nrcki.ru/}. Calculations of the molecular spectrum of LuOH$^+$ are supported by the Russian Science Foundation grant no. 24-12-00092 (\url{https://rscf.ru/project/24-12-00092/}).
The compact basis set construction was supported by the Foundation for the Advancement of Theoretical Physics and Mathematics ‘BASIS’ Grant according to Project No. 24-1-1-36-1.

\end{acknowledgments}
\section*{Author Declarations}

\subsection*{Conflict of interest}
The authors have no conflicts to disclose.

\section*{Availability of data}
The data that support the findings of this study are available from the corresponding author upon reasonable request.


\begin{thebibliography}{41}%
\makeatletter
\providecommand \@ifxundefined [1]{%
 \@ifx{#1\undefined}
}%
\providecommand \@ifnum [1]{%
 \ifnum #1\expandafter \@firstoftwo
 \else \expandafter \@secondoftwo
 \fi
}%
\providecommand \@ifx [1]{%
 \ifx #1\expandafter \@firstoftwo
 \else \expandafter \@secondoftwo
 \fi
}%
\providecommand \natexlab [1]{#1}%
\providecommand \enquote  [1]{``#1''}%
\providecommand \bibnamefont  [1]{#1}%
\providecommand \bibfnamefont [1]{#1}%
\providecommand \citenamefont [1]{#1}%
\providecommand \href@noop [0]{\@secondoftwo}%
\providecommand \href [0]{\begingroup \@sanitize@url \@href}%
\providecommand \@href[1]{\@@startlink{#1}\@@href}%
\providecommand \@@href[1]{\endgroup#1\@@endlink}%
\providecommand \@sanitize@url [0]{\catcode `\\12\catcode `\$12\catcode `\&12\catcode `\#12\catcode `\^12\catcode `\_12\catcode `\%12\relax}%
\providecommand \@@startlink[1]{}%
\providecommand \@@endlink[0]{}%
\providecommand \url  [0]{\begingroup\@sanitize@url \@url }%
\providecommand \@url [1]{\endgroup\@href {#1}{\urlprefix }}%
\providecommand \urlprefix  [0]{URL }%
\providecommand \Eprint [0]{\href }%
\providecommand \doibase [0]{https://doi.org/}%
\providecommand \selectlanguage [0]{\@gobble}%
\providecommand \bibinfo  [0]{\@secondoftwo}%
\providecommand \bibfield  [0]{\@secondoftwo}%
\providecommand \translation [1]{[#1]}%
\providecommand \BibitemOpen [0]{}%
\providecommand \bibitemStop [0]{}%
\providecommand \bibitemNoStop [0]{.\EOS\space}%
\providecommand \EOS [0]{\spacefactor3000\relax}%
\providecommand \BibitemShut  [1]{\csname bibitem#1\endcsname}%
\let\auto@bib@innerbib\@empty
\bibitem [{\citenamefont {Schwartz}(2014)}]{schwartz2014quantum}%
  \BibitemOpen
  \bibfield  {author} {\bibinfo {author} {\bibfnamefont {M.~D.}\ \bibnamefont {Schwartz}},\ }\href@noop {} {\emph {\bibinfo {title} {Quantum field theory and the standard model}}}\ (\bibinfo  {publisher} {Cambridge University Press},\ \bibinfo {year} {2014})\BibitemShut {NoStop}%
\bibitem [{\citenamefont {{Particle Data Group}}\ \emph {et~al.}(2020)\citenamefont {{Particle Data Group}}, \citenamefont {Zyla}, \citenamefont {Barnett}, \citenamefont {Beringer}, \citenamefont {Dahl}, \citenamefont {Dwyer}, \citenamefont {Groom}, \citenamefont {Lin}, \citenamefont {Lugovsky}, \citenamefont {Pianori} \emph {et~al.}}]{particle2020review}%
  \BibitemOpen
  \bibfield  {author} {\bibinfo {author} {\bibnamefont {{Particle Data Group}}}, \bibinfo {author} {\bibfnamefont {P.}~\bibnamefont {Zyla}}, \bibinfo {author} {\bibfnamefont {R.}~\bibnamefont {Barnett}}, \bibinfo {author} {\bibfnamefont {J.}~\bibnamefont {Beringer}}, \bibinfo {author} {\bibfnamefont {O.}~\bibnamefont {Dahl}}, \bibinfo {author} {\bibfnamefont {D.}~\bibnamefont {Dwyer}}, \bibinfo {author} {\bibfnamefont {D.}~\bibnamefont {Groom}}, \bibinfo {author} {\bibfnamefont {C.-J.}\ \bibnamefont {Lin}}, \bibinfo {author} {\bibfnamefont {K.}~\bibnamefont {Lugovsky}}, \bibinfo {author} {\bibfnamefont {E.}~\bibnamefont {Pianori}}, \emph {et~al.},\ }\bibfield  {title} {\enquote {\bibinfo {title} {Review of particle physics},}\ }\href@noop {} {\bibfield  {journal} {\bibinfo  {journal} {Progress of Theoretical and Experimental Physics}\ }\textbf {\bibinfo {volume} {2020}},\ \bibinfo {pages} {083C01} (\bibinfo {year} {2020})}\BibitemShut {NoStop}%
\bibitem [{\citenamefont {Khriplovich}\ and\ \citenamefont {Lamoreaux}(2012)}]{khriplovich2012cp}%
  \BibitemOpen
  \bibfield  {author} {\bibinfo {author} {\bibfnamefont {I.~B.}\ \bibnamefont {Khriplovich}}\ and\ \bibinfo {author} {\bibfnamefont {S.~K.}\ \bibnamefont {Lamoreaux}},\ }\href@noop {} {\emph {\bibinfo {title} {CP violation without strangeness: electric dipole moments of particles, atoms, and molecules}}}\ (\bibinfo  {publisher} {Springer Science \& Business Media},\ \bibinfo {year} {2012})\BibitemShut {NoStop}%
\bibitem [{\citenamefont {Cabibbo}(1963)}]{Cabibbo1963}%
  \BibitemOpen
  \bibfield  {author} {\bibinfo {author} {\bibfnamefont {N.}~\bibnamefont {Cabibbo}},\ }\bibfield  {title} {\enquote {\bibinfo {title} {{Unitary Symmetry and Leptonic Decays}},}\ }\href {https://doi.org/10.1103/PhysRevLett.10.531} {\bibfield  {journal} {\bibinfo  {journal} {Phys. Rev. Lett.}\ }\textbf {\bibinfo {volume} {10}},\ \bibinfo {pages} {531--533} (\bibinfo {year} {1963})}\BibitemShut {NoStop}%
\bibitem [{\citenamefont {Kobayashi}\ and\ \citenamefont {Maskawa}(1973)}]{KobayashiMaskawa1973}%
  \BibitemOpen
  \bibfield  {author} {\bibinfo {author} {\bibfnamefont {M.}~\bibnamefont {Kobayashi}}\ and\ \bibinfo {author} {\bibfnamefont {T.}~\bibnamefont {Maskawa}},\ }\bibfield  {title} {\enquote {\bibinfo {title} {{CP Violation in the Renormalizable Theory of Weak Interaction}},}\ }\href {https://doi.org/10.1143/PTP.49.652} {\bibfield  {journal} {\bibinfo  {journal} {Prog. Theor. Phys.}\ }\textbf {\bibinfo {volume} {49}},\ \bibinfo {pages} {652--657} (\bibinfo {year} {1973})}\BibitemShut {NoStop}%
\bibitem [{\citenamefont {Pontecorvo}(1957)}]{Pontecorvo1957}%
  \BibitemOpen
  \bibfield  {author} {\bibinfo {author} {\bibfnamefont {B.}~\bibnamefont {Pontecorvo}},\ }\bibfield  {title} {\enquote {\bibinfo {title} {{Inverse beta processes and nonconservation of lepton charge}},}\ }\href@noop {} {\bibfield  {journal} {\bibinfo  {journal} {Zh. Eksp. Teor. Fiz.}\ }\textbf {\bibinfo {volume} {34}},\ \bibinfo {pages} {247} (\bibinfo {year} {1957})}\BibitemShut {NoStop}%
\bibitem [{\citenamefont {Maki}, \citenamefont {Nakagawa},\ and\ \citenamefont {Sakata}(1962)}]{MNS1962}%
  \BibitemOpen
  \bibfield  {author} {\bibinfo {author} {\bibfnamefont {Z.}~\bibnamefont {Maki}}, \bibinfo {author} {\bibfnamefont {M.}~\bibnamefont {Nakagawa}},\ and\ \bibinfo {author} {\bibfnamefont {S.}~\bibnamefont {Sakata}},\ }\bibfield  {title} {\enquote {\bibinfo {title} {{Remarks on the unified model of elementary particles}},}\ }\href {https://doi.org/10.1143/PTP.28.870} {\bibfield  {journal} {\bibinfo  {journal} {Prog. Theor. Phys.}\ }\textbf {\bibinfo {volume} {28}},\ \bibinfo {pages} {870--880} (\bibinfo {year} {1962})}\BibitemShut {NoStop}%
\bibitem [{\citenamefont {Aaij}(2019)}]{CharmDecay19}%
  \BibitemOpen
  \bibfield  {author} {\bibinfo {author} {\bibfnamefont {R.}~\bibnamefont {Aaij}} (\bibinfo {collaboration} {LHCb Collaboration}),\ }\bibfield  {title} {\enquote {\bibinfo {title} {Observation of $cp$ violation in charm decays},}\ }\href {https://doi.org/10.1103/PhysRevLett.122.211803} {\bibfield  {journal} {\bibinfo  {journal} {Phys. Rev. Lett.}\ }\textbf {\bibinfo {volume} {122}},\ \bibinfo {pages} {211803} (\bibinfo {year} {2019})}\BibitemShut {NoStop}%
\bibitem [{\citenamefont {Pospelov}\ and\ \citenamefont {Ritz}(2014)}]{PospelovRitz2014}%
  \BibitemOpen
  \bibfield  {author} {\bibinfo {author} {\bibfnamefont {M.}~\bibnamefont {Pospelov}}\ and\ \bibinfo {author} {\bibfnamefont {A.}~\bibnamefont {Ritz}},\ }\bibfield  {title} {\enquote {\bibinfo {title} {{CKM benchmarks for electron electric dipole moment experiments}},}\ }\href {https://doi.org/10.1103/PhysRevD.89.056006} {\bibfield  {journal} {\bibinfo  {journal} {Phys. Rev. D}\ }\textbf {\bibinfo {volume} {89}},\ \bibinfo {pages} {056006} (\bibinfo {year} {2014})},\ \Eprint {https://arxiv.org/abs/1311.5537} {arXiv:1311.5537 [hep-ph]} \BibitemShut {NoStop}%
\bibitem [{\citenamefont {Yamaguchi}\ and\ \citenamefont {Yamanaka}(2020)}]{YamaguchiYamanaka2020}%
  \BibitemOpen
  \bibfield  {author} {\bibinfo {author} {\bibfnamefont {Y.}~\bibnamefont {Yamaguchi}}\ and\ \bibinfo {author} {\bibfnamefont {N.}~\bibnamefont {Yamanaka}},\ }\bibfield  {title} {\enquote {\bibinfo {title} {{Large long-distance contributions to the electric dipole moments of charged leptons in the standard model}},}\ }\href {https://doi.org/10.1103/PhysRevLett.125.241802} {\bibfield  {journal} {\bibinfo  {journal} {Phys. Rev. Lett.}\ }\textbf {\bibinfo {volume} {125}},\ \bibinfo {pages} {241802} (\bibinfo {year} {2020})},\ \Eprint {https://arxiv.org/abs/2003.08195} {arXiv:2003.08195 [hep-ph]} \BibitemShut {NoStop}%
\bibitem [{\citenamefont {FUKUYAMA}(2012)}]{Fukuyama2012}%
  \BibitemOpen
  \bibfield  {author} {\bibinfo {author} {\bibfnamefont {T.}~\bibnamefont {FUKUYAMA}},\ }\bibfield  {title} {\enquote {\bibinfo {title} {{SEARCHING FOR NEW PHYSICS BEYOND THE STANDARD MODEL IN ELECTRIC DIPOLE MOMENT}},}\ }\href {https://doi.org/10.1142/S0217751X12300153} {\bibfield  {journal} {\bibinfo  {journal} {International Journal of Modern Physics A}\ }\textbf {\bibinfo {volume} {27}},\ \bibinfo {pages} {1230015} (\bibinfo {year} {2012})}\BibitemShut {NoStop}%
\bibitem [{\citenamefont {Yamaguchi}\ and\ \citenamefont {Yamanaka}(2021)}]{YamaguchiYamanaka2021}%
  \BibitemOpen
  \bibfield  {author} {\bibinfo {author} {\bibfnamefont {Y.}~\bibnamefont {Yamaguchi}}\ and\ \bibinfo {author} {\bibfnamefont {N.}~\bibnamefont {Yamanaka}},\ }\bibfield  {title} {\enquote {\bibinfo {title} {{Quark level and hadronic contributions to the electric dipole moment of charged leptons in the standard model}},}\ }\href {https://doi.org/10.1103/PhysRevD.103.013001} {\bibfield  {journal} {\bibinfo  {journal} {Phys. Rev. D}\ }\textbf {\bibinfo {volume} {103}},\ \bibinfo {pages} {013001} (\bibinfo {year} {2021})},\ \Eprint {https://arxiv.org/abs/2006.00281} {arXiv:2006.00281 [hep-ph]} \BibitemShut {NoStop}%
\bibitem [{\citenamefont {Alarcon}\ \emph {et~al.}(2022)\citenamefont {Alarcon}, \citenamefont {Alexander}, \citenamefont {Anastassopoulos}, \citenamefont {Aoki}, \citenamefont {Baartman}, \citenamefont {Baeßler}, \citenamefont {Bartoszek}, \citenamefont {Beck}, \citenamefont {Bedeschi}, \citenamefont {Berger}, \citenamefont {Berz}, \citenamefont {Bethlem}, \citenamefont {Bhattacharya}, \citenamefont {Blaskiewicz}, \citenamefont {Blum}, \citenamefont {Bowcock}, \citenamefont {Borschevsky}, \citenamefont {Brown}, \citenamefont {Budker}, \citenamefont {Burdin}, \citenamefont {Casey}, \citenamefont {Casse}, \citenamefont {Cantatore}, \citenamefont {Cheng}, \citenamefont {Chupp}, \citenamefont {Cianciolo}, \citenamefont {Cirigliano}, \citenamefont {Clayton}, \citenamefont {Crawford}, \citenamefont {Das}, \citenamefont {Davoudiasl}, \citenamefont {de~Vries}, \citenamefont {DeMille}, \citenamefont {Denisov}, \citenamefont {Diwan}, \citenamefont {Doyle}, \citenamefont {Engel}, \citenamefont {Fanourakis},
  \citenamefont {Fatemi}, \citenamefont {Filippone}, \citenamefont {Flambaum}, \citenamefont {Fleig}, \citenamefont {Fomin}, \citenamefont {Fischer}, \citenamefont {Gabrielse}, \citenamefont {Ruiz}, \citenamefont {Gardikiotis}, \citenamefont {Gatti}, \citenamefont {Geraci}, \citenamefont {Gooding}, \citenamefont {Golub}, \citenamefont {Graham}, \citenamefont {Gray}, \citenamefont {Griffith}, \citenamefont {Haciomeroglu}, \citenamefont {Gwinner}, \citenamefont {Hoekstra}, \citenamefont {Hoffstaetter}, \citenamefont {Huang}, \citenamefont {Hutzler}, \citenamefont {Incagli}, \citenamefont {Ito}, \citenamefont {Izubuchi}, \citenamefont {Jayich}, \citenamefont {Jeong}, \citenamefont {Kaplan}, \citenamefont {Karuza}, \citenamefont {Kawall}, \citenamefont {Kim}, \citenamefont {Koop}, \citenamefont {Korsch}, \citenamefont {Korobkina}, \citenamefont {Lebedev}, \citenamefont {Lee}, \citenamefont {Lee}, \citenamefont {Lehnert}, \citenamefont {Leung}, \citenamefont {Liu}, \citenamefont {Long}, \citenamefont {Lusiani},
  \citenamefont {Marciano}, \citenamefont {Maroudas}, \citenamefont {Matlashov}, \citenamefont {Matsumoto}, \citenamefont {Mawhorter}, \citenamefont {Meot}, \citenamefont {Mereghetti}, \citenamefont {Miller}, \citenamefont {Morse}, \citenamefont {Mott}, \citenamefont {Omarov}, \citenamefont {Orozco}, \citenamefont {O'Shaughnessy}, \citenamefont {Ozben}, \citenamefont {Park}, \citenamefont {Pattie}, \citenamefont {Petrov}, \citenamefont {Piacentino}, \citenamefont {Plaster}, \citenamefont {Podobedov}, \citenamefont {Poelker}, \citenamefont {Pocanic}, \citenamefont {Prasannaa}, \citenamefont {Price}, \citenamefont {Ramsey-Musolf}, \citenamefont {Raparia}, \citenamefont {Rajendran}, \citenamefont {Reece}, \citenamefont {Reid}, \citenamefont {Rescia}, \citenamefont {Ritz}, \citenamefont {Roberts}, \citenamefont {Safronova}, \citenamefont {Sakemi}, \citenamefont {Schmidt-Wellenburg}, \citenamefont {Shindler}, \citenamefont {Semertzidis}, \citenamefont {Silenko}, \citenamefont {Singh}, \citenamefont {Skripnikov},
  \citenamefont {Soni}, \citenamefont {Stephenson}, \citenamefont {Suleiman}, \citenamefont {Sunaga}, \citenamefont {Syphers}, \citenamefont {Syritsyn}, \citenamefont {Tarbutt}, \citenamefont {Thoerngren}, \citenamefont {Timmermans}, \citenamefont {Tishchenko}, \citenamefont {Titov}, \citenamefont {Tsoupas}, \citenamefont {Tzamarias}, \citenamefont {Variola}, \citenamefont {Venanzoni}, \citenamefont {Vilella}, \citenamefont {Vossebeld}, \citenamefont {Winter}, \citenamefont {Won}, \citenamefont {Zelenski}, \citenamefont {Zelevinsky}, \citenamefont {Zhou},\ and\ \citenamefont {Zioutas}}]{whitepaper}%
  \BibitemOpen
  \bibfield  {author} {\bibinfo {author} {\bibfnamefont {R.}~\bibnamefont {Alarcon}}, \bibinfo {author} {\bibfnamefont {J.}~\bibnamefont {Alexander}}, \bibinfo {author} {\bibfnamefont {V.}~\bibnamefont {Anastassopoulos}}, \bibinfo {author} {\bibfnamefont {T.}~\bibnamefont {Aoki}}, \bibinfo {author} {\bibfnamefont {R.}~\bibnamefont {Baartman}}, \bibinfo {author} {\bibfnamefont {S.}~\bibnamefont {Baeßler}}, \bibinfo {author} {\bibfnamefont {L.}~\bibnamefont {Bartoszek}}, \bibinfo {author} {\bibfnamefont {D.~H.}\ \bibnamefont {Beck}}, \bibinfo {author} {\bibfnamefont {F.}~\bibnamefont {Bedeschi}}, \bibinfo {author} {\bibfnamefont {R.}~\bibnamefont {Berger}}, \bibinfo {author} {\bibfnamefont {M.}~\bibnamefont {Berz}}, \bibinfo {author} {\bibfnamefont {H.~L.}\ \bibnamefont {Bethlem}}, \bibinfo {author} {\bibfnamefont {T.}~\bibnamefont {Bhattacharya}}, \bibinfo {author} {\bibfnamefont {M.}~\bibnamefont {Blaskiewicz}}, \bibinfo {author} {\bibfnamefont {T.}~\bibnamefont {Blum}}, \bibinfo {author} {\bibfnamefont
  {T.}~\bibnamefont {Bowcock}}, \bibinfo {author} {\bibfnamefont {A.}~\bibnamefont {Borschevsky}}, \bibinfo {author} {\bibfnamefont {K.}~\bibnamefont {Brown}}, \bibinfo {author} {\bibfnamefont {D.}~\bibnamefont {Budker}}, \bibinfo {author} {\bibfnamefont {S.}~\bibnamefont {Burdin}}, \bibinfo {author} {\bibfnamefont {B.~C.}\ \bibnamefont {Casey}}, \bibinfo {author} {\bibfnamefont {G.}~\bibnamefont {Casse}}, \bibinfo {author} {\bibfnamefont {G.}~\bibnamefont {Cantatore}}, \bibinfo {author} {\bibfnamefont {L.}~\bibnamefont {Cheng}}, \bibinfo {author} {\bibfnamefont {T.}~\bibnamefont {Chupp}}, \bibinfo {author} {\bibfnamefont {V.}~\bibnamefont {Cianciolo}}, \bibinfo {author} {\bibfnamefont {V.}~\bibnamefont {Cirigliano}}, \bibinfo {author} {\bibfnamefont {S.~M.}\ \bibnamefont {Clayton}}, \bibinfo {author} {\bibfnamefont {C.}~\bibnamefont {Crawford}}, \bibinfo {author} {\bibfnamefont {B.~P.}\ \bibnamefont {Das}}, \bibinfo {author} {\bibfnamefont {H.}~\bibnamefont {Davoudiasl}}, \bibinfo {author} {\bibfnamefont
  {J.}~\bibnamefont {de~Vries}}, \bibinfo {author} {\bibfnamefont {D.}~\bibnamefont {DeMille}}, \bibinfo {author} {\bibfnamefont {D.}~\bibnamefont {Denisov}}, \bibinfo {author} {\bibfnamefont {M.~V.}\ \bibnamefont {Diwan}}, \bibinfo {author} {\bibfnamefont {J.~M.}\ \bibnamefont {Doyle}}, \bibinfo {author} {\bibfnamefont {J.}~\bibnamefont {Engel}}, \bibinfo {author} {\bibfnamefont {G.}~\bibnamefont {Fanourakis}}, \bibinfo {author} {\bibfnamefont {R.}~\bibnamefont {Fatemi}}, \bibinfo {author} {\bibfnamefont {B.~W.}\ \bibnamefont {Filippone}}, \bibinfo {author} {\bibfnamefont {V.~V.}\ \bibnamefont {Flambaum}}, \bibinfo {author} {\bibfnamefont {T.}~\bibnamefont {Fleig}}, \bibinfo {author} {\bibfnamefont {N.}~\bibnamefont {Fomin}}, \bibinfo {author} {\bibfnamefont {W.}~\bibnamefont {Fischer}}, \bibinfo {author} {\bibfnamefont {G.}~\bibnamefont {Gabrielse}}, \bibinfo {author} {\bibfnamefont {R.~F.~G.}\ \bibnamefont {Ruiz}}, \bibinfo {author} {\bibfnamefont {A.}~\bibnamefont {Gardikiotis}}, \bibinfo {author}
  {\bibfnamefont {C.}~\bibnamefont {Gatti}}, \bibinfo {author} {\bibfnamefont {A.}~\bibnamefont {Geraci}}, \bibinfo {author} {\bibfnamefont {J.}~\bibnamefont {Gooding}}, \bibinfo {author} {\bibfnamefont {B.}~\bibnamefont {Golub}}, \bibinfo {author} {\bibfnamefont {P.}~\bibnamefont {Graham}}, \bibinfo {author} {\bibfnamefont {F.}~\bibnamefont {Gray}}, \bibinfo {author} {\bibfnamefont {W.~C.}\ \bibnamefont {Griffith}}, \bibinfo {author} {\bibfnamefont {S.}~\bibnamefont {Haciomeroglu}}, \bibinfo {author} {\bibfnamefont {G.}~\bibnamefont {Gwinner}}, \bibinfo {author} {\bibfnamefont {S.}~\bibnamefont {Hoekstra}}, \bibinfo {author} {\bibfnamefont {G.~H.}\ \bibnamefont {Hoffstaetter}}, \bibinfo {author} {\bibfnamefont {H.}~\bibnamefont {Huang}}, \bibinfo {author} {\bibfnamefont {N.~R.}\ \bibnamefont {Hutzler}}, \bibinfo {author} {\bibfnamefont {M.}~\bibnamefont {Incagli}}, \bibinfo {author} {\bibfnamefont {T.~M.}\ \bibnamefont {Ito}}, \bibinfo {author} {\bibfnamefont {T.}~\bibnamefont {Izubuchi}}, \bibinfo {author}
  {\bibfnamefont {A.~M.}\ \bibnamefont {Jayich}}, \bibinfo {author} {\bibfnamefont {H.}~\bibnamefont {Jeong}}, \bibinfo {author} {\bibfnamefont {D.}~\bibnamefont {Kaplan}}, \bibinfo {author} {\bibfnamefont {M.}~\bibnamefont {Karuza}}, \bibinfo {author} {\bibfnamefont {D.}~\bibnamefont {Kawall}}, \bibinfo {author} {\bibfnamefont {O.}~\bibnamefont {Kim}}, \bibinfo {author} {\bibfnamefont {I.}~\bibnamefont {Koop}}, \bibinfo {author} {\bibfnamefont {W.}~\bibnamefont {Korsch}}, \bibinfo {author} {\bibfnamefont {E.}~\bibnamefont {Korobkina}}, \bibinfo {author} {\bibfnamefont {V.}~\bibnamefont {Lebedev}}, \bibinfo {author} {\bibfnamefont {J.}~\bibnamefont {Lee}}, \bibinfo {author} {\bibfnamefont {S.}~\bibnamefont {Lee}}, \bibinfo {author} {\bibfnamefont {R.}~\bibnamefont {Lehnert}}, \bibinfo {author} {\bibfnamefont {K.~K.~H.}\ \bibnamefont {Leung}}, \bibinfo {author} {\bibfnamefont {C.-Y.}\ \bibnamefont {Liu}}, \bibinfo {author} {\bibfnamefont {J.}~\bibnamefont {Long}}, \bibinfo {author} {\bibfnamefont
  {A.}~\bibnamefont {Lusiani}}, \bibinfo {author} {\bibfnamefont {W.~J.}\ \bibnamefont {Marciano}}, \bibinfo {author} {\bibfnamefont {M.}~\bibnamefont {Maroudas}}, \bibinfo {author} {\bibfnamefont {A.}~\bibnamefont {Matlashov}}, \bibinfo {author} {\bibfnamefont {N.}~\bibnamefont {Matsumoto}}, \bibinfo {author} {\bibfnamefont {R.}~\bibnamefont {Mawhorter}}, \bibinfo {author} {\bibfnamefont {F.}~\bibnamefont {Meot}}, \bibinfo {author} {\bibfnamefont {E.}~\bibnamefont {Mereghetti}}, \bibinfo {author} {\bibfnamefont {J.~P.}\ \bibnamefont {Miller}}, \bibinfo {author} {\bibfnamefont {W.~M.}\ \bibnamefont {Morse}}, \bibinfo {author} {\bibfnamefont {J.}~\bibnamefont {Mott}}, \bibinfo {author} {\bibfnamefont {Z.}~\bibnamefont {Omarov}}, \bibinfo {author} {\bibfnamefont {L.~A.}\ \bibnamefont {Orozco}}, \bibinfo {author} {\bibfnamefont {C.~M.}\ \bibnamefont {O'Shaughnessy}}, \bibinfo {author} {\bibfnamefont {C.}~\bibnamefont {Ozben}}, \bibinfo {author} {\bibfnamefont {S.}~\bibnamefont {Park}}, \bibinfo {author}
  {\bibfnamefont {R.~W.}\ \bibnamefont {Pattie}}, \bibinfo {author} {\bibfnamefont {A.~N.}\ \bibnamefont {Petrov}}, \bibinfo {author} {\bibfnamefont {G.~M.}\ \bibnamefont {Piacentino}}, \bibinfo {author} {\bibfnamefont {B.~R.}\ \bibnamefont {Plaster}}, \bibinfo {author} {\bibfnamefont {B.}~\bibnamefont {Podobedov}}, \bibinfo {author} {\bibfnamefont {M.}~\bibnamefont {Poelker}}, \bibinfo {author} {\bibfnamefont {D.}~\bibnamefont {Pocanic}}, \bibinfo {author} {\bibfnamefont {V.~S.}\ \bibnamefont {Prasannaa}}, \bibinfo {author} {\bibfnamefont {J.}~\bibnamefont {Price}}, \bibinfo {author} {\bibfnamefont {M.~J.}\ \bibnamefont {Ramsey-Musolf}}, \bibinfo {author} {\bibfnamefont {D.}~\bibnamefont {Raparia}}, \bibinfo {author} {\bibfnamefont {S.}~\bibnamefont {Rajendran}}, \bibinfo {author} {\bibfnamefont {M.}~\bibnamefont {Reece}}, \bibinfo {author} {\bibfnamefont {A.}~\bibnamefont {Reid}}, \bibinfo {author} {\bibfnamefont {S.}~\bibnamefont {Rescia}}, \bibinfo {author} {\bibfnamefont {A.}~\bibnamefont {Ritz}},
  \bibinfo {author} {\bibfnamefont {B.~L.}\ \bibnamefont {Roberts}}, \bibinfo {author} {\bibfnamefont {M.~S.}\ \bibnamefont {Safronova}}, \bibinfo {author} {\bibfnamefont {Y.}~\bibnamefont {Sakemi}}, \bibinfo {author} {\bibfnamefont {P.}~\bibnamefont {Schmidt-Wellenburg}}, \bibinfo {author} {\bibfnamefont {A.}~\bibnamefont {Shindler}}, \bibinfo {author} {\bibfnamefont {Y.~K.}\ \bibnamefont {Semertzidis}}, \bibinfo {author} {\bibfnamefont {A.}~\bibnamefont {Silenko}}, \bibinfo {author} {\bibfnamefont {J.~T.}\ \bibnamefont {Singh}}, \bibinfo {author} {\bibfnamefont {L.~V.}\ \bibnamefont {Skripnikov}}, \bibinfo {author} {\bibfnamefont {A.}~\bibnamefont {Soni}}, \bibinfo {author} {\bibfnamefont {E.}~\bibnamefont {Stephenson}}, \bibinfo {author} {\bibfnamefont {R.}~\bibnamefont {Suleiman}}, \bibinfo {author} {\bibfnamefont {A.}~\bibnamefont {Sunaga}}, \bibinfo {author} {\bibfnamefont {M.}~\bibnamefont {Syphers}}, \bibinfo {author} {\bibfnamefont {S.}~\bibnamefont {Syritsyn}}, \bibinfo {author} {\bibfnamefont
  {M.~R.}\ \bibnamefont {Tarbutt}}, \bibinfo {author} {\bibfnamefont {P.}~\bibnamefont {Thoerngren}}, \bibinfo {author} {\bibfnamefont {R.~G.~E.}\ \bibnamefont {Timmermans}}, \bibinfo {author} {\bibfnamefont {V.}~\bibnamefont {Tishchenko}}, \bibinfo {author} {\bibfnamefont {A.~V.}\ \bibnamefont {Titov}}, \bibinfo {author} {\bibfnamefont {N.}~\bibnamefont {Tsoupas}}, \bibinfo {author} {\bibfnamefont {S.}~\bibnamefont {Tzamarias}}, \bibinfo {author} {\bibfnamefont {A.}~\bibnamefont {Variola}}, \bibinfo {author} {\bibfnamefont {G.}~\bibnamefont {Venanzoni}}, \bibinfo {author} {\bibfnamefont {E.}~\bibnamefont {Vilella}}, \bibinfo {author} {\bibfnamefont {J.}~\bibnamefont {Vossebeld}}, \bibinfo {author} {\bibfnamefont {P.}~\bibnamefont {Winter}}, \bibinfo {author} {\bibfnamefont {E.}~\bibnamefont {Won}}, \bibinfo {author} {\bibfnamefont {A.}~\bibnamefont {Zelenski}}, \bibinfo {author} {\bibfnamefont {T.}~\bibnamefont {Zelevinsky}}, \bibinfo {author} {\bibfnamefont {Y.}~\bibnamefont {Zhou}},\ and\ \bibinfo {author}
  {\bibfnamefont {K.}~\bibnamefont {Zioutas}},\ }\href {https://doi.org/10.48550/ARXIV.2203.08103} {\enquote {\bibinfo {title} {Electric dipole moments and the search for new physics},}\ } (\bibinfo {year} {2022})\BibitemShut {NoStop}%
\bibitem [{\citenamefont {Roussy}\ \emph {et~al.}(2023)\citenamefont {Roussy}, \citenamefont {Caldwell}, \citenamefont {Wright}, \citenamefont {Cairncross}, \citenamefont {Shagam}, \citenamefont {Ng}, \citenamefont {Schlossberger}, \citenamefont {Park}, \citenamefont {Wang}, \citenamefont {Ye},\ and\ \citenamefont {Cornell}}]{JILA23}%
  \BibitemOpen
  \bibfield  {author} {\bibinfo {author} {\bibfnamefont {T.~S.}\ \bibnamefont {Roussy}}, \bibinfo {author} {\bibfnamefont {L.}~\bibnamefont {Caldwell}}, \bibinfo {author} {\bibfnamefont {T.}~\bibnamefont {Wright}}, \bibinfo {author} {\bibfnamefont {W.~B.}\ \bibnamefont {Cairncross}}, \bibinfo {author} {\bibfnamefont {Y.}~\bibnamefont {Shagam}}, \bibinfo {author} {\bibfnamefont {K.~B.}\ \bibnamefont {Ng}}, \bibinfo {author} {\bibfnamefont {N.}~\bibnamefont {Schlossberger}}, \bibinfo {author} {\bibfnamefont {S.~Y.}\ \bibnamefont {Park}}, \bibinfo {author} {\bibfnamefont {A.}~\bibnamefont {Wang}}, \bibinfo {author} {\bibfnamefont {J.}~\bibnamefont {Ye}},\ and\ \bibinfo {author} {\bibfnamefont {E.~A.}\ \bibnamefont {Cornell}},\ }\bibfield  {title} {\enquote {\bibinfo {title} {An improved bound on the electron’s electric dipole moment},}\ }\href {https://doi.org/10.1126/science.adg4084} {\bibfield  {journal} {\bibinfo  {journal} {Science}\ }\textbf {\bibinfo {volume} {381}},\ \bibinfo {pages} {46--50} (\bibinfo
  {year} {2023})},\ \Eprint {https://arxiv.org/abs/https://www.science.org/doi/pdf/10.1126/science.adg4084} {https://www.science.org/doi/pdf/10.1126/science.adg4084} \BibitemShut {NoStop}%
\bibitem [{\citenamefont {Kim}\ and\ \citenamefont {Carosi}(2010)}]{KimCarosi2010}%
  \BibitemOpen
  \bibfield  {author} {\bibinfo {author} {\bibfnamefont {J.~E.}\ \bibnamefont {Kim}}\ and\ \bibinfo {author} {\bibfnamefont {G.}~\bibnamefont {Carosi}},\ }\bibfield  {title} {\enquote {\bibinfo {title} {{Axions and the Strong CP Problem}},}\ }\href {https://doi.org/10.1103/RevModPhys.82.557} {\bibfield  {journal} {\bibinfo  {journal} {Rev. Mod. Phys.}\ }\textbf {\bibinfo {volume} {82}},\ \bibinfo {pages} {557--602} (\bibinfo {year} {2010})},\ \bibinfo {note} {[Erratum: Rev.Mod.Phys. 91, 049902 (2019)]},\ \Eprint {https://arxiv.org/abs/0807.3125} {arXiv:0807.3125 [hep-ph]} \BibitemShut {NoStop}%
\bibitem [{\citenamefont {Mulder}, \citenamefont {Timmermans},\ and\ \citenamefont {de~Vries}(2025)}]{Mulder2025}%
  \BibitemOpen
  \bibfield  {author} {\bibinfo {author} {\bibfnamefont {H.}~\bibnamefont {Mulder}}, \bibinfo {author} {\bibfnamefont {R.}~\bibnamefont {Timmermans}},\ and\ \bibinfo {author} {\bibfnamefont {J.}~\bibnamefont {de~Vries}},\ }\bibfield  {title} {\enquote {\bibinfo {title} {Probing the qcd $\theta$ term with paramagnetic molecules},}\ }\href {https://doi.org/10.1007/JHEP07(2025)232} {\bibfield  {journal} {\bibinfo  {journal} {Journal of High Energy Physics}\ } (\bibinfo {year} {2025})}\BibitemShut {NoStop}%
\bibitem [{\citenamefont {Isaev}\ and\ \citenamefont {Berger}(2016)}]{Isaev:16}%
  \BibitemOpen
  \bibfield  {author} {\bibinfo {author} {\bibfnamefont {T.~A.}\ \bibnamefont {Isaev}}\ and\ \bibinfo {author} {\bibfnamefont {R.}~\bibnamefont {Berger}},\ }\bibfield  {title} {\enquote {\bibinfo {title} {Polyatomic candidates for cooling of molecules with lasers from simple theoretical concepts},}\ }\href {https://doi.org/10.1103/PhysRevLett.116.063006} {\bibfield  {journal} {\bibinfo  {journal} {Phys. Rev. Lett.}\ }\textbf {\bibinfo {volume} {116}},\ \bibinfo {pages} {063006} (\bibinfo {year} {2016})}\BibitemShut {NoStop}%
\bibitem [{\citenamefont {Maison}\ \emph {et~al.}(2020)\citenamefont {Maison}, \citenamefont {Skripnikov}, \citenamefont {Flambaum},\ and\ \citenamefont {Grau}}]{Maison:20a}%
  \BibitemOpen
  \bibfield  {author} {\bibinfo {author} {\bibfnamefont {D.~E.}\ \bibnamefont {Maison}}, \bibinfo {author} {\bibfnamefont {L.~V.}\ \bibnamefont {Skripnikov}}, \bibinfo {author} {\bibfnamefont {V.~V.}\ \bibnamefont {Flambaum}},\ and\ \bibinfo {author} {\bibfnamefont {M.}~\bibnamefont {Grau}},\ }\bibfield  {title} {\enquote {\bibinfo {title} {Search for $\mathcal{CP}$-violating nuclear magnetic quadrupole moment using the {LuOH$^+$} cation},}\ }\href {https://doi.org/10.1063/5.0028983} {\bibfield  {journal} {\bibinfo  {journal} {J. Chem. Phys.}\ }\textbf {\bibinfo {volume} {153}},\ \bibinfo {pages} {224302} (\bibinfo {year} {2020})}\BibitemShut {NoStop}%
\bibitem [{\citenamefont {Kozyryev}\ and\ \citenamefont {Hutzler}(2017)}]{Kozyryev:17}%
  \BibitemOpen
  \bibfield  {author} {\bibinfo {author} {\bibfnamefont {I.}~\bibnamefont {Kozyryev}}\ and\ \bibinfo {author} {\bibfnamefont {N.~R.}\ \bibnamefont {Hutzler}},\ }\bibfield  {title} {\enquote {\bibinfo {title} {Precision measurement of time-reversal symmetry violation with laser-cooled polyatomic molecules},}\ }\href {https://doi.org/10.1103/PhysRevLett.119.133002} {\bibfield  {journal} {\bibinfo  {journal} {Phys. Rev. Lett.}\ }\textbf {\bibinfo {volume} {119}},\ \bibinfo {pages} {133002} (\bibinfo {year} {2017})}\BibitemShut {NoStop}%
\bibitem [{\citenamefont {Hutzler}(2020)}]{hutzler2020polyatomic}%
  \BibitemOpen
  \bibfield  {author} {\bibinfo {author} {\bibfnamefont {N.~R.}\ \bibnamefont {Hutzler}},\ }\bibfield  {title} {\enquote {\bibinfo {title} {Polyatomic molecules as quantum sensors for fundamental physics},}\ }\href@noop {} {\bibfield  {journal} {\bibinfo  {journal} {Quantum Science and Technology}\ }\textbf {\bibinfo {volume} {5}},\ \bibinfo {pages} {044011} (\bibinfo {year} {2020})}\BibitemShut {NoStop}%
\bibitem [{\citenamefont {Petrov}\ and\ \citenamefont {Zakharova}(2022)}]{Petrov:2022}%
  \BibitemOpen
  \bibfield  {author} {\bibinfo {author} {\bibfnamefont {A.}~\bibnamefont {Petrov}}\ and\ \bibinfo {author} {\bibfnamefont {A.}~\bibnamefont {Zakharova}},\ }\bibfield  {title} {\enquote {\bibinfo {title} {Sensitivity of the yboh molecule to $\mathcal{P}\mathcal{T}$-odd effects in an external electric field},}\ }\href {https://doi.org/10.1103/PhysRevA.105.L050801} {\bibfield  {journal} {\bibinfo  {journal} {Phys. Rev. A}\ }\textbf {\bibinfo {volume} {105}},\ \bibinfo {pages} {L050801} (\bibinfo {year} {2022})}\BibitemShut {NoStop}%
\bibitem [{\citenamefont {Petrov}(2024{\natexlab{a}})}]{Petrov:24b}%
  \BibitemOpen
  \bibfield  {author} {\bibinfo {author} {\bibfnamefont {A.}~\bibnamefont {Petrov}},\ }\bibfield  {title} {\enquote {\bibinfo {title} {Electric-field-dependent $g$ factors of a yboh molecule},}\ }\href {https://doi.org/10.1103/PhysRevA.110.L030804} {\bibfield  {journal} {\bibinfo  {journal} {Phys. Rev. A}\ }\textbf {\bibinfo {volume} {110}},\ \bibinfo {pages} {L030804} (\bibinfo {year} {2024}{\natexlab{a}})}\BibitemShut {NoStop}%
\bibitem [{\citenamefont {Petrov}(2024{\natexlab{b}})}]{Petrov:24a}%
  \BibitemOpen
  \bibfield  {author} {\bibinfo {author} {\bibfnamefont {A.}~\bibnamefont {Petrov}},\ }\bibfield  {title} {\enquote {\bibinfo {title} {Electronic matrix elements for parity doubling in the yboh molecule},}\ }\href {https://doi.org/10.1103/PhysRevA.109.012819} {\bibfield  {journal} {\bibinfo  {journal} {Phys. Rev. A}\ }\textbf {\bibinfo {volume} {109}},\ \bibinfo {pages} {012819} (\bibinfo {year} {2024}{\natexlab{b}})}\BibitemShut {NoStop}%
\bibitem [{\citenamefont {Zakharova}\ and\ \citenamefont {Petrov}(2022)}]{zakharova22}%
  \BibitemOpen
  \bibfield  {author} {\bibinfo {author} {\bibfnamefont {A.}~\bibnamefont {Zakharova}}\ and\ \bibinfo {author} {\bibfnamefont {A.}~\bibnamefont {Petrov}},\ }\bibfield  {title} {\enquote {\bibinfo {title} {Impact of ligand deformation on the p,t-violation effects in the yboh molecule},}\ }\href {https://doi.org/10.1063/5.0121110} {\bibfield  {journal} {\bibinfo  {journal} {The Journal of Chemical Physics}\ }\textbf {\bibinfo {volume} {157}},\ \bibinfo {pages} {154310} (\bibinfo {year} {2022})},\ \Eprint {https://arxiv.org/abs/https://pubs.aip.org/aip/jcp/article-pdf/doi/10.1063/5.0121110/16552613/154310\_1\_online.pdf} {https://pubs.aip.org/aip/jcp/article-pdf/doi/10.1063/5.0121110/16552613/154310\_1\_online.pdf} \BibitemShut {NoStop}%
\bibitem [{\citenamefont {Jadbabaie}\ \emph {et~al.}(2023)\citenamefont {Jadbabaie}, \citenamefont {Takahashi}, \citenamefont {Pilgram}, \citenamefont {Conn}, \citenamefont {Zeng}, \citenamefont {Zhang},\ and\ \citenamefont {Hutzler}}]{jadbabaie23}%
  \BibitemOpen
  \bibfield  {author} {\bibinfo {author} {\bibfnamefont {A.}~\bibnamefont {Jadbabaie}}, \bibinfo {author} {\bibfnamefont {Y.}~\bibnamefont {Takahashi}}, \bibinfo {author} {\bibfnamefont {N.~H.}\ \bibnamefont {Pilgram}}, \bibinfo {author} {\bibfnamefont {C.~J.}\ \bibnamefont {Conn}}, \bibinfo {author} {\bibfnamefont {Y.}~\bibnamefont {Zeng}}, \bibinfo {author} {\bibfnamefont {C.}~\bibnamefont {Zhang}},\ and\ \bibinfo {author} {\bibfnamefont {N.~R.}\ \bibnamefont {Hutzler}},\ }\bibfield  {title} {\enquote {\bibinfo {title} {Characterizing the fundamental bending vibration of a linear polyatomic molecule for symmetry violation searches},}\ }\href {https://doi.org/10.1088/1367-2630/ace471} {\bibfield  {journal} {\bibinfo  {journal} {New J. Phys. 073014}\ }\textbf {\bibinfo {volume} {25}},\ \bibinfo {pages} {073014} (\bibinfo {year} {2023})}\BibitemShut {NoStop}%
\bibitem [{\citenamefont {Kurchavov}\ \emph {et~al.}(2023)\citenamefont {Kurchavov}, \citenamefont {Maison}, \citenamefont {Skripnikov}, \citenamefont {Grau},\ and\ \citenamefont {Petrov}}]{kurchavov23}%
  \BibitemOpen
  \bibfield  {author} {\bibinfo {author} {\bibfnamefont {I.}~\bibnamefont {Kurchavov}}, \bibinfo {author} {\bibfnamefont {D.}~\bibnamefont {Maison}}, \bibinfo {author} {\bibfnamefont {L.}~\bibnamefont {Skripnikov}}, \bibinfo {author} {\bibfnamefont {M.}~\bibnamefont {Grau}},\ and\ \bibinfo {author} {\bibfnamefont {A.}~\bibnamefont {Petrov}},\ }\bibfield  {title} {\enquote {\bibinfo {title} {Nuclear magnetic quadrupole moment of $^{175}\mathrm{Lu}$ and parity-violating polarization degree of levels in $^{175}\mathrm{Lu}{\text{oh}}^{+}$},}\ }\href {https://doi.org/10.1103/PhysRevA.108.052815} {\bibfield  {journal} {\bibinfo  {journal} {Phys. Rev. A}\ }\textbf {\bibinfo {volume} {108}},\ \bibinfo {pages} {052815} (\bibinfo {year} {2023})}\BibitemShut {NoStop}%
\bibitem [{\citenamefont {McGuire}\ and\ \citenamefont {Kouri}(1974)}]{mcguire1974quantum}%
  \BibitemOpen
  \bibfield  {author} {\bibinfo {author} {\bibfnamefont {P.}~\bibnamefont {McGuire}}\ and\ \bibinfo {author} {\bibfnamefont {D.~J.}\ \bibnamefont {Kouri}},\ }\bibfield  {title} {\enquote {\bibinfo {title} {{Quantum mechanical close coupling approach to molecular collisions. $j_z$-conserving coupled states approximation}},}\ }\href@noop {} {\bibfield  {journal} {\bibinfo  {journal} {The Journal of Chemical Physics}\ }\textbf {\bibinfo {volume} {60}},\ \bibinfo {pages} {2488--2499} (\bibinfo {year} {1974})}\BibitemShut {NoStop}%
\bibitem [{\citenamefont {Gomes}, \citenamefont {Dyall},\ and\ \citenamefont {Visscher}(2010)}]{gomes2010relativistic}%
  \BibitemOpen
  \bibfield  {author} {\bibinfo {author} {\bibfnamefont {A.~S.}\ \bibnamefont {Gomes}}, \bibinfo {author} {\bibfnamefont {K.~G.}\ \bibnamefont {Dyall}},\ and\ \bibinfo {author} {\bibfnamefont {L.}~\bibnamefont {Visscher}},\ }\bibfield  {title} {\enquote {\bibinfo {title} {Relativistic double-zeta, triple-zeta, and quadruple-zeta basis sets for the lanthanides la--lu},}\ }\href@noop {} {\bibfield  {journal} {\bibinfo  {journal} {Theoretical Chemistry Accounts}\ }\textbf {\bibinfo {volume} {127}},\ \bibinfo {pages} {369--381} (\bibinfo {year} {2010})}\BibitemShut {NoStop}%
\bibitem [{\citenamefont {{Dunning, Jr}}(1989)}]{Dunning:89}%
  \BibitemOpen
  \bibfield  {author} {\bibinfo {author} {\bibfnamefont {T.~H.}\ \bibnamefont {{Dunning, Jr}}},\ }\bibfield  {title} {\enquote {\bibinfo {title} {Gaussian basis sets for use in correlated molecular calculations. {I}. {T}he atoms boron through neon and hydrogen},}\ }\href {https://doi.org/10.1063/1.456153} {\bibfield  {journal} {\bibinfo  {journal} {J. Comp. Phys.}\ }\textbf {\bibinfo {volume} {90}},\ \bibinfo {pages} {1007--1023} (\bibinfo {year} {1989})}\BibitemShut {NoStop}%
\bibitem [{\citenamefont {Kendall}, \citenamefont {{Dunning, Jr}},\ and\ \citenamefont {Harrison}(1992)}]{Kendall:92}%
  \BibitemOpen
  \bibfield  {author} {\bibinfo {author} {\bibfnamefont {R.~A.}\ \bibnamefont {Kendall}}, \bibinfo {author} {\bibfnamefont {T.~H.}\ \bibnamefont {{Dunning, Jr}}},\ and\ \bibinfo {author} {\bibfnamefont {R.~J.}\ \bibnamefont {Harrison}},\ }\bibfield  {title} {\enquote {\bibinfo {title} {Electron affinities of the first-row atoms revisited. {S}ystematic basis sets and wave functions},}\ }\href {https://doi.org/10.1063/1.462569} {\bibfield  {journal} {\bibinfo  {journal} {J. Comp. Phys.}\ }\textbf {\bibinfo {volume} {96}},\ \bibinfo {pages} {6796--6806} (\bibinfo {year} {1992})}\BibitemShut {NoStop}%
\bibitem [{\citenamefont {Titov}\ and\ \citenamefont {Mosyagin}(1999)}]{titov1999generalized}%
  \BibitemOpen
  \bibfield  {author} {\bibinfo {author} {\bibfnamefont {A.}~\bibnamefont {Titov}}\ and\ \bibinfo {author} {\bibfnamefont {N.}~\bibnamefont {Mosyagin}},\ }\bibfield  {title} {\enquote {\bibinfo {title} {Generalized relativistic effective core potential: Theoretical grounds},}\ }\href@noop {} {\bibfield  {journal} {\bibinfo  {journal} {International journal of quantum chemistry}\ }\textbf {\bibinfo {volume} {71}},\ \bibinfo {pages} {359--401} (\bibinfo {year} {1999})}\BibitemShut {NoStop}%
\bibitem [{\citenamefont {Mosyagin}, \citenamefont {Zaitsevskii},\ and\ \citenamefont {Titov}(2010)}]{mosyagin2010shape}%
  \BibitemOpen
  \bibfield  {author} {\bibinfo {author} {\bibfnamefont {N.~S.}\ \bibnamefont {Mosyagin}}, \bibinfo {author} {\bibfnamefont {A.}~\bibnamefont {Zaitsevskii}},\ and\ \bibinfo {author} {\bibfnamefont {A.~V.}\ \bibnamefont {Titov}},\ }\bibfield  {title} {\enquote {\bibinfo {title} {Shape-consistent relativistic effective potentials of small atomic cores},}\ }\href@noop {} {\bibfield  {journal} {\bibinfo  {journal} {International Review of Atomic and Molecular Physics}\ }\textbf {\bibinfo {volume} {1}},\ \bibinfo {pages} {63--72} (\bibinfo {year} {2010})}\BibitemShut {NoStop}%
\bibitem [{\citenamefont {Mosyagin}\ \emph {et~al.}(2016)\citenamefont {Mosyagin}, \citenamefont {Zaitsevskii}, \citenamefont {Skripnikov},\ and\ \citenamefont {Titov}}]{mosyagin2016generalized}%
  \BibitemOpen
  \bibfield  {author} {\bibinfo {author} {\bibfnamefont {N.~S.}\ \bibnamefont {Mosyagin}}, \bibinfo {author} {\bibfnamefont {A.~V.}\ \bibnamefont {Zaitsevskii}}, \bibinfo {author} {\bibfnamefont {L.~V.}\ \bibnamefont {Skripnikov}},\ and\ \bibinfo {author} {\bibfnamefont {A.~V.}\ \bibnamefont {Titov}},\ }\bibfield  {title} {\enquote {\bibinfo {title} {Generalized relativistic effective core potentials for actinides},}\ }\href@noop {} {\bibfield  {journal} {\bibinfo  {journal} {International Journal of Quantum Chemistry}\ }\textbf {\bibinfo {volume} {116}},\ \bibinfo {pages} {301--315} (\bibinfo {year} {2016})}\BibitemShut {NoStop}%
\bibitem [{\citenamefont {Skripnikov}(2020)}]{Skripnikov:2020e}%
  \BibitemOpen
  \bibfield  {author} {\bibinfo {author} {\bibfnamefont {L.~V.}\ \bibnamefont {Skripnikov}},\ }\bibfield  {title} {\enquote {\bibinfo {title} {Nuclear magnetization distribution effect in molecules: Ra$^+$ and raf hyperfine structure},}\ }\href@noop {} {\bibfield  {journal} {\bibinfo  {journal} {J. Comp. Phys.}\ }\textbf {\bibinfo {volume} {153}},\ \bibinfo {pages} {114114} (\bibinfo {year} {2020})}\BibitemShut {NoStop}%
\bibitem [{\citenamefont {Skripnikov}, \citenamefont {Mosyagin},\ and\ \citenamefont {Titov}(2013)}]{Skripnikov:13a}%
  \BibitemOpen
  \bibfield  {author} {\bibinfo {author} {\bibfnamefont {L.~V.}\ \bibnamefont {Skripnikov}}, \bibinfo {author} {\bibfnamefont {N.~S.}\ \bibnamefont {Mosyagin}},\ and\ \bibinfo {author} {\bibfnamefont {A.~V.}\ \bibnamefont {Titov}},\ }\bibfield  {title} {\enquote {\bibinfo {title} {Relativistic coupled-cluster calculations of spectroscopic and chemical properties for element 120},}\ }\href@noop {} {\ \textbf {\bibinfo {volume} {555}},\ \bibinfo {pages} {79--83} (\bibinfo {year} {2013})}\BibitemShut {NoStop}%
\bibitem [{\citenamefont {Bartlett}\ and\ \citenamefont {Musia{\l}}(2007)}]{Bartlett:2007}%
  \BibitemOpen
  \bibfield  {author} {\bibinfo {author} {\bibfnamefont {R.~J.}\ \bibnamefont {Bartlett}}\ and\ \bibinfo {author} {\bibfnamefont {M.}~\bibnamefont {Musia{\l}}},\ }\bibfield  {title} {\enquote {\bibinfo {title} {Coupled-cluster theory in quantum chemistry},}\ }\href {https://doi.org/10.1103/RevModPhys.79.291} {\bibfield  {journal} {\bibinfo  {journal} {Rev. Mod. Phys.}\ }\textbf {\bibinfo {volume} {79}},\ \bibinfo {pages} {291--352} (\bibinfo {year} {2007})}\BibitemShut {NoStop}%
\bibitem [{DIR()}]{DIRAC19}%
  \BibitemOpen
  \href@noop {} {}\bibinfo {note} {{DIRAC}, a relativistic ab initio electronic structure program, Release {DIRAC19} (2019), written by A.~S.~P.~Gomes, T.~Saue, L.~Visscher, H.~J.~{\relax Aa}.~Jensen, and R.~Bast, with contributions from I.~A.~Aucar, V.~Bakken, K.~G.~Dyall, S.~Dubillard, U.~Ekstr{\"o}m, E.~Eliav, T.~Enevoldsen, E.~Fa{\ss}hauer, T.~Fleig, O.~Fossgaard, L.~Halbert, E.~D.~Hedeg{\aa}rd, B.~Heimlich--Paris, T.~Helgaker, J.~Henriksson, M.~Ilia{\v{s}}, Ch.~R.~Jacob, S.~Knecht, S.~Komorovsk{\'y}, O.~Kullie, J.~K.~L{\ae}rdahl, C.~V.~Larsen, Y.~S.~Lee, H.~S.~Nataraj, M.~K.~Nayak, P.~Norman, G.~Olejniczak, J.~Olsen, J.~M.~H.~Olsen, Y.~C.~Park, J.~K.~Pedersen, M.~Pernpointner, R.~di~Remigio, K.~Ruud, P.~Sa{\l}ek, B.~Schimmelpfennig, B.~Senjean, A.~Shee, J.~Sikkema, A.~J.~Thorvaldsen, J.~Thyssen, J.~van~Stralen, M.~L.~Vidal, S.~Villaume, O.~Visser, T.~Winther, and S.~Yamamoto (available at http://dx.doi.org/10.5281/zenodo.3572669, see also http://www.diracprogram.org)}\BibitemShut {NoStop}%
\bibitem [{\citenamefont {Saue}\ \emph {et~al.}(2020)\citenamefont {Saue}, \citenamefont {Bast}, \citenamefont {Gomes}, \citenamefont {Jensen}, \citenamefont {Visscher}, \citenamefont {Aucar}, \citenamefont {Di~Remigio}, \citenamefont {Dyall}, \citenamefont {Eliav}, \citenamefont {Fasshauer}, \citenamefont {Fleig}, \citenamefont {Halbert}, \citenamefont {Hedegard}, \citenamefont {Helmich-Paris}, \citenamefont {Ilias}, \citenamefont {Jacob}, \citenamefont {Knecht}, \citenamefont {Laerdahl}, \citenamefont {Vidal}, \citenamefont {Nayak}, \citenamefont {Olejniczak}, \citenamefont {Olsen}, \citenamefont {Pernpointner}, \citenamefont {Senjean}, \citenamefont {Shee}, \citenamefont {Sunaga},\ and\ \citenamefont {van Stralen}}]{Saue:2020}%
  \BibitemOpen
  \bibfield  {author} {\bibinfo {author} {\bibfnamefont {T.}~\bibnamefont {Saue}}, \bibinfo {author} {\bibfnamefont {R.}~\bibnamefont {Bast}}, \bibinfo {author} {\bibfnamefont {A.~S.~P.}\ \bibnamefont {Gomes}}, \bibinfo {author} {\bibfnamefont {H.~J.~A.}\ \bibnamefont {Jensen}}, \bibinfo {author} {\bibfnamefont {L.}~\bibnamefont {Visscher}}, \bibinfo {author} {\bibfnamefont {I.~A.}\ \bibnamefont {Aucar}}, \bibinfo {author} {\bibfnamefont {R.}~\bibnamefont {Di~Remigio}}, \bibinfo {author} {\bibfnamefont {K.~G.}\ \bibnamefont {Dyall}}, \bibinfo {author} {\bibfnamefont {E.}~\bibnamefont {Eliav}}, \bibinfo {author} {\bibfnamefont {E.}~\bibnamefont {Fasshauer}}, \bibinfo {author} {\bibfnamefont {T.}~\bibnamefont {Fleig}}, \bibinfo {author} {\bibfnamefont {L.}~\bibnamefont {Halbert}}, \bibinfo {author} {\bibfnamefont {E.~D.}\ \bibnamefont {Hedegard}}, \bibinfo {author} {\bibfnamefont {B.}~\bibnamefont {Helmich-Paris}}, \bibinfo {author} {\bibfnamefont {M.}~\bibnamefont {Ilias}}, \bibinfo {author} {\bibfnamefont
  {C.~R.}\ \bibnamefont {Jacob}}, \bibinfo {author} {\bibfnamefont {S.}~\bibnamefont {Knecht}}, \bibinfo {author} {\bibfnamefont {J.~K.}\ \bibnamefont {Laerdahl}}, \bibinfo {author} {\bibfnamefont {M.~L.}\ \bibnamefont {Vidal}}, \bibinfo {author} {\bibfnamefont {M.~K.}\ \bibnamefont {Nayak}}, \bibinfo {author} {\bibfnamefont {M.}~\bibnamefont {Olejniczak}}, \bibinfo {author} {\bibfnamefont {J.~M.~H.}\ \bibnamefont {Olsen}}, \bibinfo {author} {\bibfnamefont {M.}~\bibnamefont {Pernpointner}}, \bibinfo {author} {\bibfnamefont {B.}~\bibnamefont {Senjean}}, \bibinfo {author} {\bibfnamefont {A.}~\bibnamefont {Shee}}, \bibinfo {author} {\bibfnamefont {A.}~\bibnamefont {Sunaga}},\ and\ \bibinfo {author} {\bibfnamefont {J.~N.~P.}\ \bibnamefont {van Stralen}},\ }\bibfield  {title} {\enquote {\bibinfo {title} {The dirac code for relativistic molecular calculations},}\ }\href {https://doi.org/10.1063/5.0004844} {\bibfield  {journal} {\bibinfo  {journal} {J. Chem. Phys.}\ }\textbf {\bibinfo {volume} {152}},\ \bibinfo
  {pages} {204104} (\bibinfo {year} {2020})}\BibitemShut {NoStop}%
\bibitem [{\citenamefont {Stanton}\ \emph {et~al.}(2011)\citenamefont {Stanton}, \citenamefont {Gauss}, \citenamefont {Harding}, \citenamefont {Szalay} \emph {et~al.}}]{CFOUR}%
  \BibitemOpen
  \bibfield  {author} {\bibinfo {author} {\bibfnamefont {J.~F.}\ \bibnamefont {Stanton}}, \bibinfo {author} {\bibfnamefont {J.}~\bibnamefont {Gauss}}, \bibinfo {author} {\bibfnamefont {M.~E.}\ \bibnamefont {Harding}}, \bibinfo {author} {\bibfnamefont {P.~G.}\ \bibnamefont {Szalay}}, \emph {et~al.},\ }\href@noop {} {\enquote {\bibinfo {title} {{``{\sc cfour}''}},}\ } (\bibinfo {year} {2011}),\ \bibinfo {note} {{\sc cfour}: a program package for performing high-level quantum chemical calculations on atoms and molecules, {http://www.cfour.de} .}\BibitemShut {Stop}%
\bibitem [{\citenamefont {Herzberg}(1966)}]{HerzbergBook}%
  \BibitemOpen
  \bibfield  {author} {\bibinfo {author} {\bibfnamefont {G.}~\bibnamefont {Herzberg}},\ }\href@noop {} {\emph {\bibinfo {title} {{Molecular spectra and molecular structure. Vol. 3: Electronic spectra and electronic structure of polyatomic molecules}}}}\ (\bibinfo  {publisher} {New York: Van Nostrand},\ \bibinfo {year} {1966})\BibitemShut {NoStop}%
\bibitem [{\citenamefont {Maison}\ \emph {et~al.}(2022)\citenamefont {Maison}, \citenamefont {Skripnikov}, \citenamefont {Penyazkov}, \citenamefont {Grau},\ and\ \citenamefont {Petrov}}]{Maison:22}%
  \BibitemOpen
  \bibfield  {author} {\bibinfo {author} {\bibfnamefont {D.~E.}\ \bibnamefont {Maison}}, \bibinfo {author} {\bibfnamefont {L.~V.}\ \bibnamefont {Skripnikov}}, \bibinfo {author} {\bibfnamefont {G.}~\bibnamefont {Penyazkov}}, \bibinfo {author} {\bibfnamefont {M.}~\bibnamefont {Grau}},\ and\ \bibinfo {author} {\bibfnamefont {A.~N.}\ \bibnamefont {Petrov}},\ }\bibfield  {title} {\enquote {\bibinfo {title} {$\mathcal{T},\mathcal{P}$-odd effects in the ${\mathrm{luoh}}^{+}$ cation},}\ }\href {https://doi.org/10.1103/PhysRevA.106.062827} {\bibfield  {journal} {\bibinfo  {journal} {Phys. Rev. A}\ }\textbf {\bibinfo {volume} {106}},\ \bibinfo {pages} {062827} (\bibinfo {year} {2022})}\BibitemShut {NoStop}%
\end{thebibliography}
%

\begin{table}
\footnotesize
\caption{
Calculated vibrational energy levels ($\mathrm{cm}^{-1}$) and $l$-doubling or $2q$ (MHz) at the CCSD(T) level.
Here, $\nu_1$ corresponds to $R$, $\nu_2$ to $\theta$, and $\nu_3$ to $r$. ``+OH'' and ``+SO'' indicate inclusion of ligand deformation and spin–orbit effects, respectively.
``4c'' denotes a four-component Dirac–Coulomb calculation, whereas ``1c'' designates the scalar-relativistic approximation applied to the outer-core and valence electrons.
Percent change is relative to the preceding value.
}
\begin{tabular}{lllr}
\hline \hline
              Parameter &        Basis &   Value &  \% change \\
\midrule
                            &              &         &             \\
$\nu_1=0, \nu_2=0, \nu_3=0$ &      4c/PBas   &       0 &             \\
                            &      1c/MBas   &       0 &             \\
                            &      1c/HBas   &       0 &             \\
                            &  1c/HBas +OH   &       0 &             \\
                            &1c/HBas +OH +SO &       0 &             \\
                            &              &         &             \\
$\nu_1=0, \nu_2=1, \nu_3=0$ &      4c/PBas   &  437.73 &             \\
                            &      1c/MBas   &  434.77 &       $-0.68$ \\
                            &      1c/HBas   &  460.19 &       $ 5.85$ \\
                            &  1c/HBas +OH   &  449.11 &       $-2.41$ \\
                            &1c/HBas +OH +SO &  442.23 &       $-1.53$ \\
                            &              &         &             \\
$\nu_1=1, \nu_2=0, \nu_3=0$ &      4c/PBas   &  746.14 &             \\
                            &      1c/MBas   &   744.1 &       $-0.27$ \\
                            &      1c/HBas   &  748.46 &       $ 0.59$ \\
                            &  1c/HBas +OH   &  747.83 &       $-0.08$ \\
                            &1c/HBas +OH +SO &  748.23 &       $ 0.05$ \\
                            &              &         &             \\
$\nu_1=0, \nu_2=2^0, \nu_3=0$ &      4c/PBas   &  866.93 &             \\
                              &      1c/MBas   &  855.46 &       $-1.32$ \\
                              &      1c/HBas   &  895.63 &       $ 4.70$ \\
                              &  1c/HBas +OH   &  872.33 &       $-2.60$ \\
                              &1c/HBas +OH +SO &   856.8 &       $-1.78$ \\
                              &              &         &             \\
$\nu_1=0, \nu_2=2^2, \nu_3=0$ &      4c/PBas   &  889.75 &             \\
                              &      1c/MBas   &  880.47 &       $-1.04$ \\
                              &      1c/HBas   &  926.19 &       $ 5.19$ \\
                              &  1c/HBas +OH   &  902.23 &       $-2.59$ \\
                              &1c/HBas +OH +SO &  887.65 &       $-1.62$ \\
                              &              &         &             \\
$\nu_1=0, \nu_2=0, \nu_3=1$   &      4c/PBas   &      na &             \\
                              &      1c/MBas   &      na &             \\
                              &      1c/HBas   &      na &             \\
                              &  1c/HBas +OH   & 3865.48 &             \\
                              &1c/HBas +OH +SO & 3839.81 &       $-0.66$ \\
                              &              &         &             \\
$l$-doubling                         &      4c/PBas   &   24.19 &             \\
($\nu_1=0, \nu_2=1, \nu_3=0$)        &      1c/MBas   &   24.44 &       $ 1.02$ \\
                                     &      1c/HBas   &    23.6 &       $-3.40$ \\
                                     &  1c/HBas +OH   &   25.27 &       $ 7.06$ \\
                                     &1c/HBas +OH +SO &   25.72 &       $ 1.76$ \\
                                     &              &         &             \\
$l$-doubling                           &      4c/PBas   &  0.0051 &             \\
($\nu_1=0, \nu_2=2^2, \nu_3=0$)        &      1c/MBas   &  0.0048 &      $ -6.08$ \\
                                       &      1c/HBas   &  0.0038 &      $-21.72$ \\
                                       &  1c/HBas +OH   &  0.0044 &      $ 16.99$ \\
                                       &1c/HBas +OH +SO &  0.0044 &      $  0.75$ \\
\hline \hline
\end{tabular}
\label{tab:VibSpectrum}
\end{table}

\begin{table}[H]
\caption{
Rotational spectrum parameters $B(\nu_1\nu_2^l\nu_3)$ (${\rm cm}^{-1}$) obtained at the CCSD(T) level.
Here, $\nu_1$ is associated with $R$, $\nu_2$ with $\theta$, and $\nu_3$ with $r$; ``+OH'' and ``+SO'' indicate inclusion of ligand deformation and spin–orbit effects, respectively.
``4c'' denotes a four-component Dirac–Coulomb calculation, whereas ``1c'' designates the scalar-relativistic approximation applied to the outer-core and valence electrons.
Percent changes are given relative to the preceding value.
}
\begin{tabular}{lllr}
\hline \hline
Parameter & Basis & Value & \% change \\
\midrule
$B(000)$                       &      4c/PBas   &  0.2868 &             \\
                               &      1c/MBas   &  0.2862 &       $-0.22$ \\
                               &      1c/HBas   &  0.2874 &       $ 0.42$ \\
                               &  1c/HBas +OH   &  0.2871 &       $-0.11$ \\
                               &1c/HBas +OH +SO &  0.2874 &       $ 0.09$ \\
                               &              &         &             \\
$B(010)$                       &      4c/PBas   &  0.2864 &             \\
                               &      1c/MBas   &  0.2857 &       $-0.25$ \\
                               &      1c/HBas   &  0.2869 &       $ 0.42$ \\
                               &  1c/HBas +OH   &  0.2866 &       $-0.12$ \\
                               &1c/HBas +OH +SO &  0.2868 &       $ 0.10$ \\
                               &              &         &             \\
$B(100)$                        &      4c/PBas   &  0.2855 &             \\
                                &      1c/MBas   &  0.2849 &       $-0.22$ \\
                                &      1c/HBas   &  0.2861 &       $ 0.42$ \\
                                &  1c/HBas +OH   &  0.2858 &       $-0.11$ \\
                                &1c/HBas +OH +SO &  0.2861 &       $ 0.09$ \\
                                &              &         &             \\
$B(02^00)$                       &      4c/PBas   &  0.2867 &             \\
                                 &      1c/MBas   &  0.2857 &       $-0.34$ \\
                                 &      1c/HBas   &  0.2869 &       $ 0.43$ \\
                                 &  1c/HBas +OH   &  0.2866 &       $-0.12$ \\
                                 &1c/HBas +OH +SO &  0.2869 &       $ 0.11$ \\
                                 &              &         &             \\
$B(02^20)$                       &      4c/PBas   &  0.2915 &             \\
                                 &      1c/MBas   &  0.2905 &       $-0.34$ \\
                                 &      1c/HBas   &  0.2918 &       $ 0.45$ \\
                                 &  1c/HBas +OH   &  0.2864 &       $-1.86$ \\
                                 &1c/HBas +OH +SO &  0.2867 &       $ 0.11$ \\
                                 &              &         &             \\
$B(001)$                       &      4c/PBas   &      na &             \\
                               &      1c/MBas   &      na &             \\
                               &      1c/HBas   &      na &             \\
                               &  1c/HBas +OH   &  0.2824 &             \\
                               &1c/HBas +OH +SO &  0.2827 &        $0.12$ \\
                               &              &         &             \\
\hline \hline
\end{tabular}
\label{tab:RotSpectrum}
\end{table}

\end{document}